\documentclass{webofc}
\usepackage[varg]{txfonts}

\usepackage{bm}
\usepackage{float}
\usepackage{subfigure}
\usepackage{lineno}

\begin{document}
%
\title{Elliptic and triangular flow of light nuclei in Au+Au collisions in the BES-II energies using the STAR detector}

\author{\firstname{Rishabh} \lastname{Sharma}\inst{1}\fnsep\thanks{\email{rishabhsharma@students.iisertirupati.ac.in}},
        \firstname{} \lastname{for the STAR Collaboration}
}

\institute{Indian Institute of Science Education and Research (IISER), Tirupati 517507, India}

\abstract{Light nuclei might be formed in heavy-ion collisions by the coalescence of produced (anti-)nucleons or transported nucleons. Due to their low binding energies, they are more likely to form at later stages of the hadronic fireball. In this proceedings, we report the transverse momentum and centrality dependence of elliptic ($v_{2}$) and triangular ($v_{3}$) flow of $d$, $t$, and $^3$He in Au+Au collisions at $\sqrt{s_{NN}}$ = 14.6 -- 54.4 GeV. The mass number scaling of $v_{2}(p_{T})$ and $v_{3}(p_{T})$ of light nuclei is discussed. We also report the comparison of $v_{2}(p_{T})$ and $v_{3}(p_{T})$ of light nuclei with a transport-plus-coalescence model calculation.}
\maketitle
\section{Introduction}
\label{intro}
High-energy heavy-ion collisions produce light nuclei in abundance. Thermal model proposes their formation near the chemical freezeout surface (CFO), however, due to their low binding energies it is unlikely that they survive at high CFO temperature \cite{andronic}. In contrast, the coalescence model suggests their formation at later stages via nucleon recombination \cite{RefCoal1,RefCoal2,RefCoal3,RefCoal4}. In the case of nucleon coalescence, the momentum space distributions of both the constituents (nucleons) and the products (light nuclei) are measurable in heavy-ion collision experiments. Therefore, studying the azimuthal anisotropy of light nuclei and comparing them with that of proton can give insights into the light nuclei production mechanism in heavy-ion collisions.

In the following sections, elliptic $(v_2$) and triangular ($v_3$) flow of $d$, $t$, and $^3$He in Au+Au collisions at $\sqrt{s_{NN}}$ = 14.6, 19.6, 27, and 54.4 GeV are discussed.

\section{Analysis details}
\label{analysis}
The data presented in this proceedings are from Au+Au collisions at $\sqrt{s_{NN}}$ = 14.6, 19.6, 27, and 54.4 GeV collected by the STAR experiment at RHIC during the second phase of the Beam Energy Scan (BES-II) program. Light nuclei are identified using the Time Projection Chamber (TPC) \cite{TPC} and the Time of Flight (TOF) \cite{TOF} detectors. TPC uses specific ionization energy loss ($dE/dx$) in a large gas volume for nuclei identification. Further, the purity of light nuclei signal is enhanced by imposing a constraint on their mass-square ($m^2$), measured using the Time of Flight (TOF) detector. 

The quantities $v_2$ and $v_3$ are the second and third order Fourier coefficients, respectively, characterizing the azimuthal distribution of the produced nuclei relative to the symmetry planes (called event planes) of the Au+Au collision. We have constructed the second ($\Psi_{2}$) and third ($\Psi_{3}$) order event plane angle using tracks reconstructed in the TPC. The $\eta$-subevent plane method is used to avoid auto-correlation \cite{flow}.

\section{Results}
\label{results}
\subsection{$\bm{v_{2}(p_T)}$ and $\bm{v_{3}(p_T)}$ of light nuclei}
\label{flow}
\begin{figure}[H]
    \centering
    \subfigure{\includegraphics[width=0.24\textwidth]{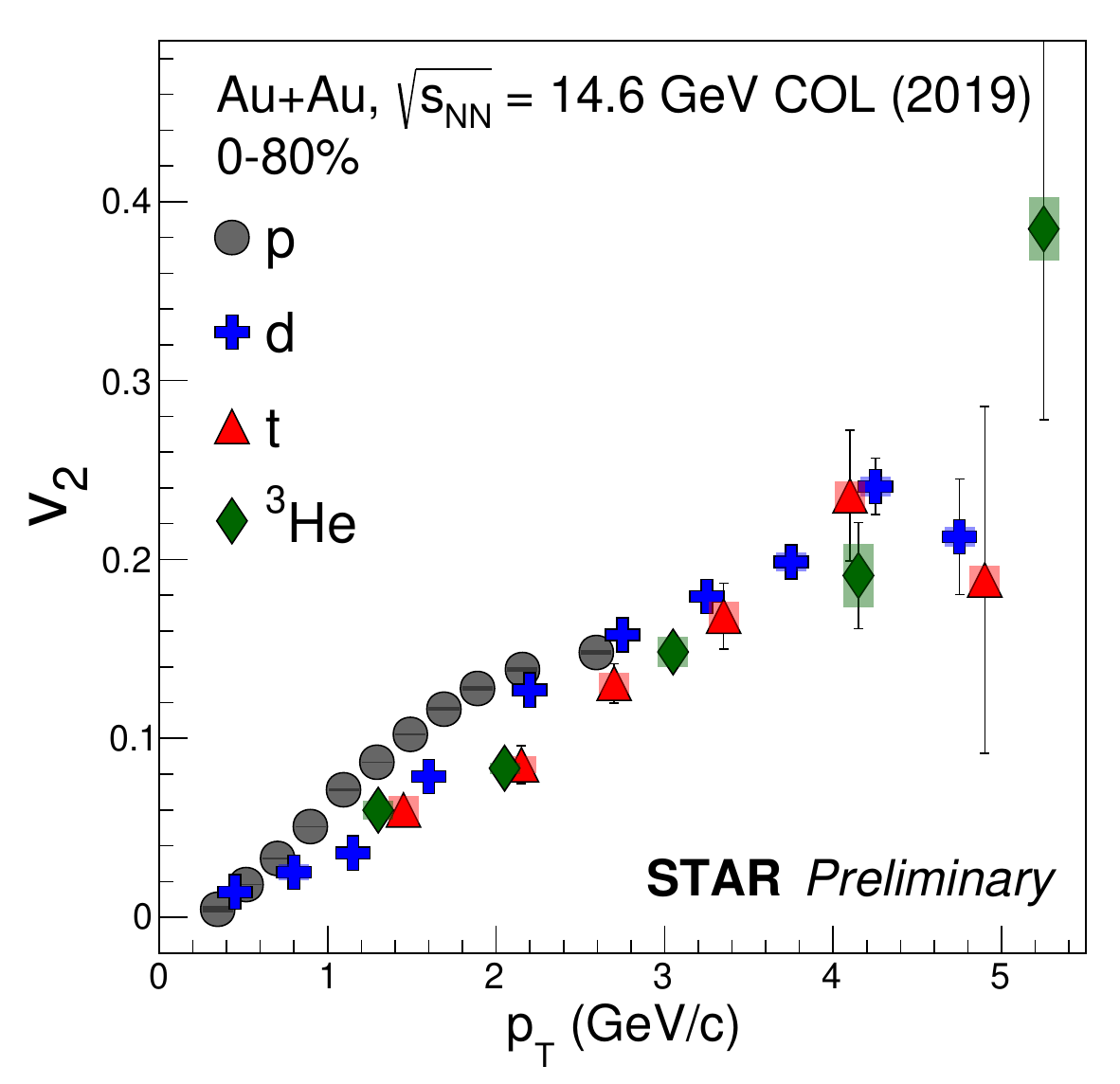}}
    \subfigure{\includegraphics[width=0.24\textwidth]{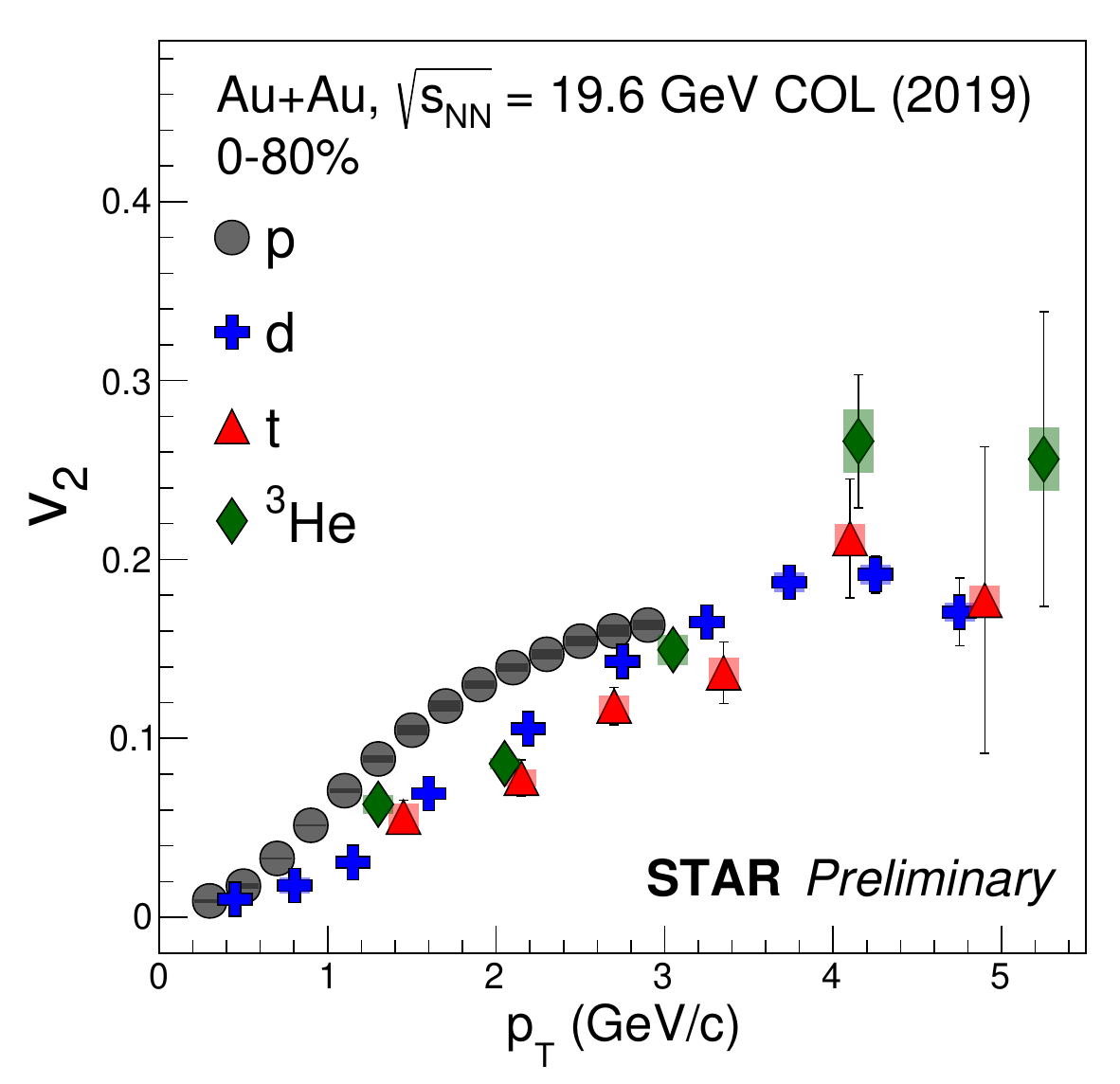}}
     \subfigure{\includegraphics[width=0.24\textwidth]{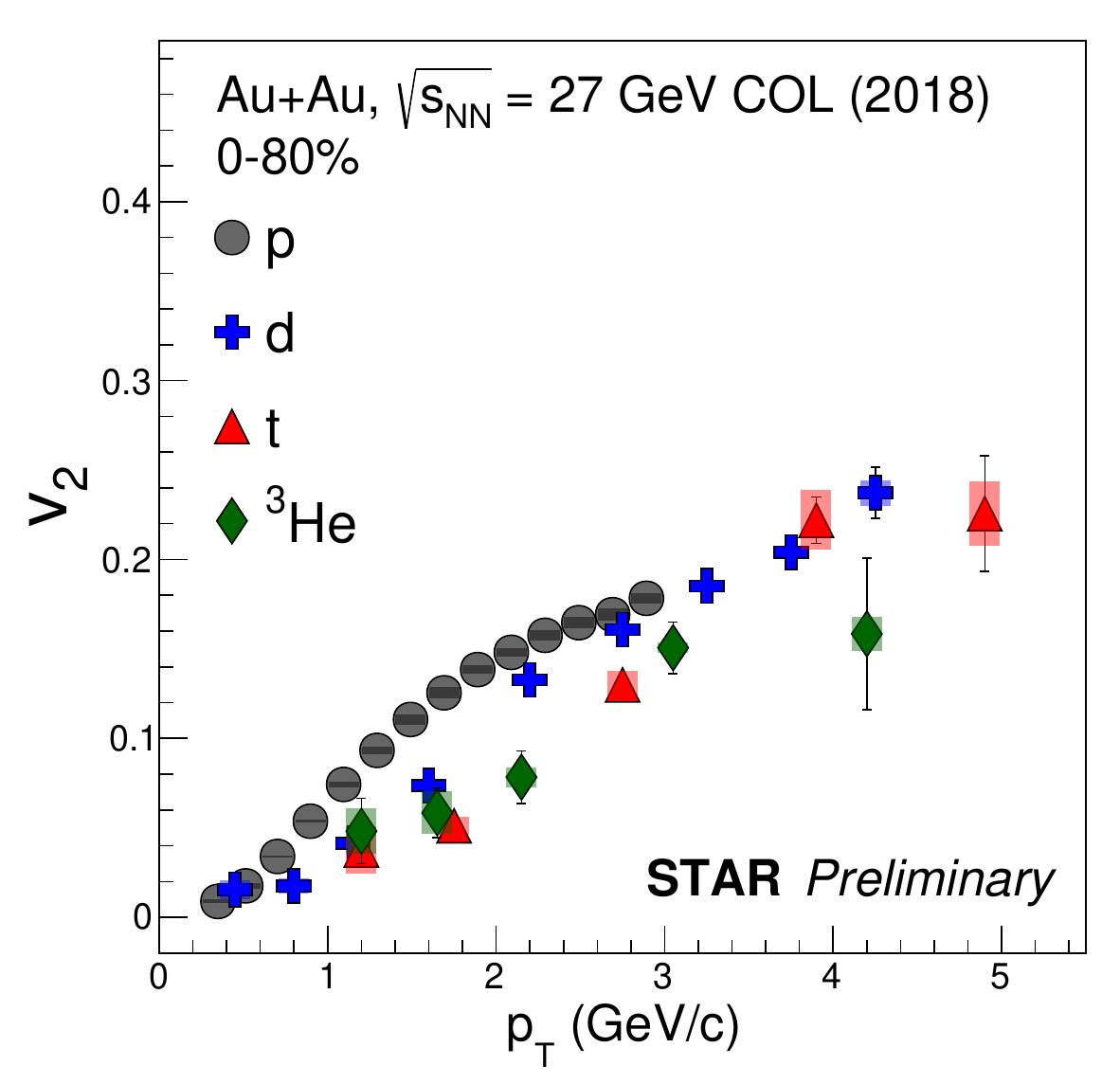}}
    \subfigure{\includegraphics[width=0.24\textwidth]{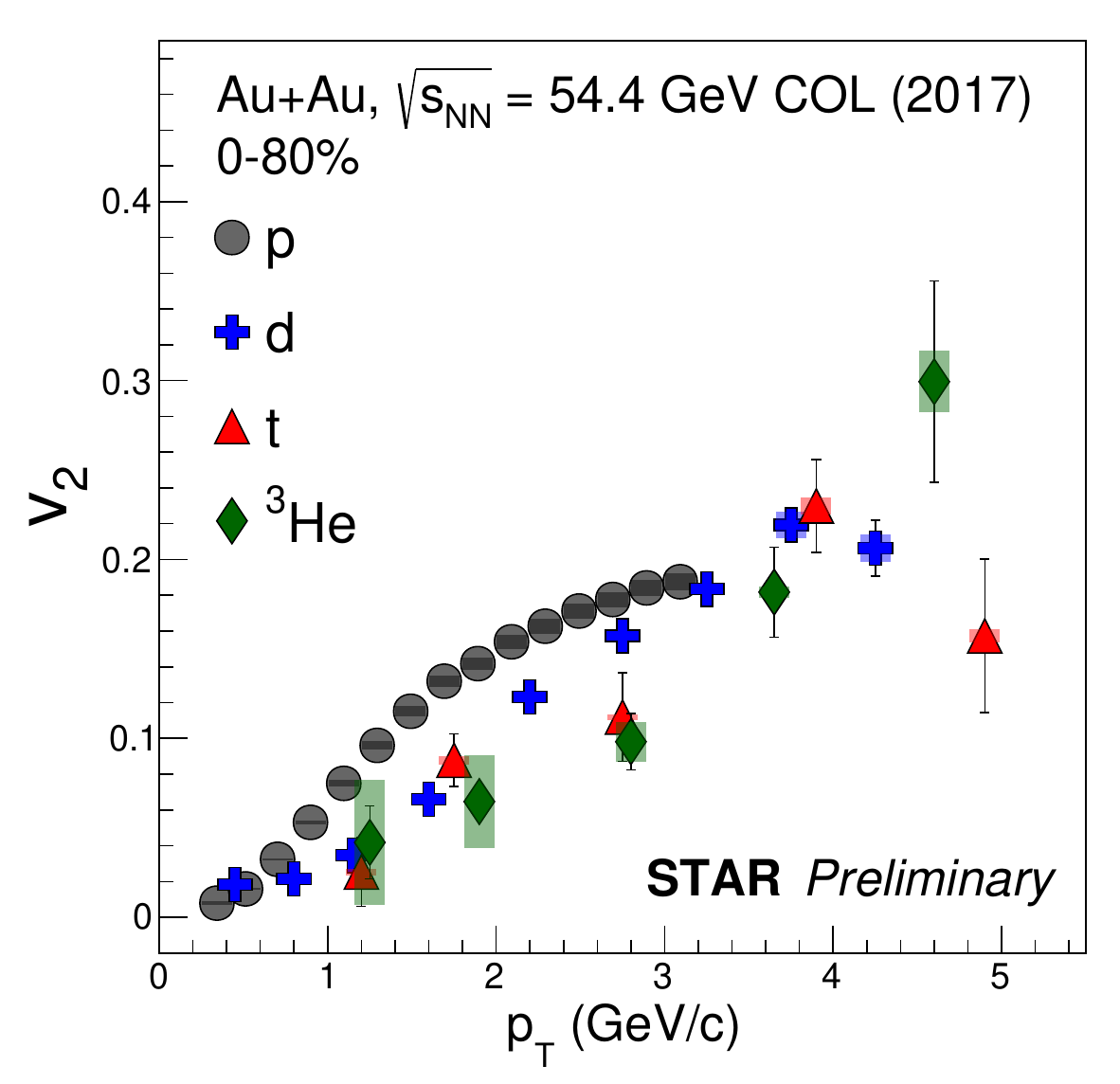}} \newline
    \subfigure{\includegraphics[width=0.24\textwidth]{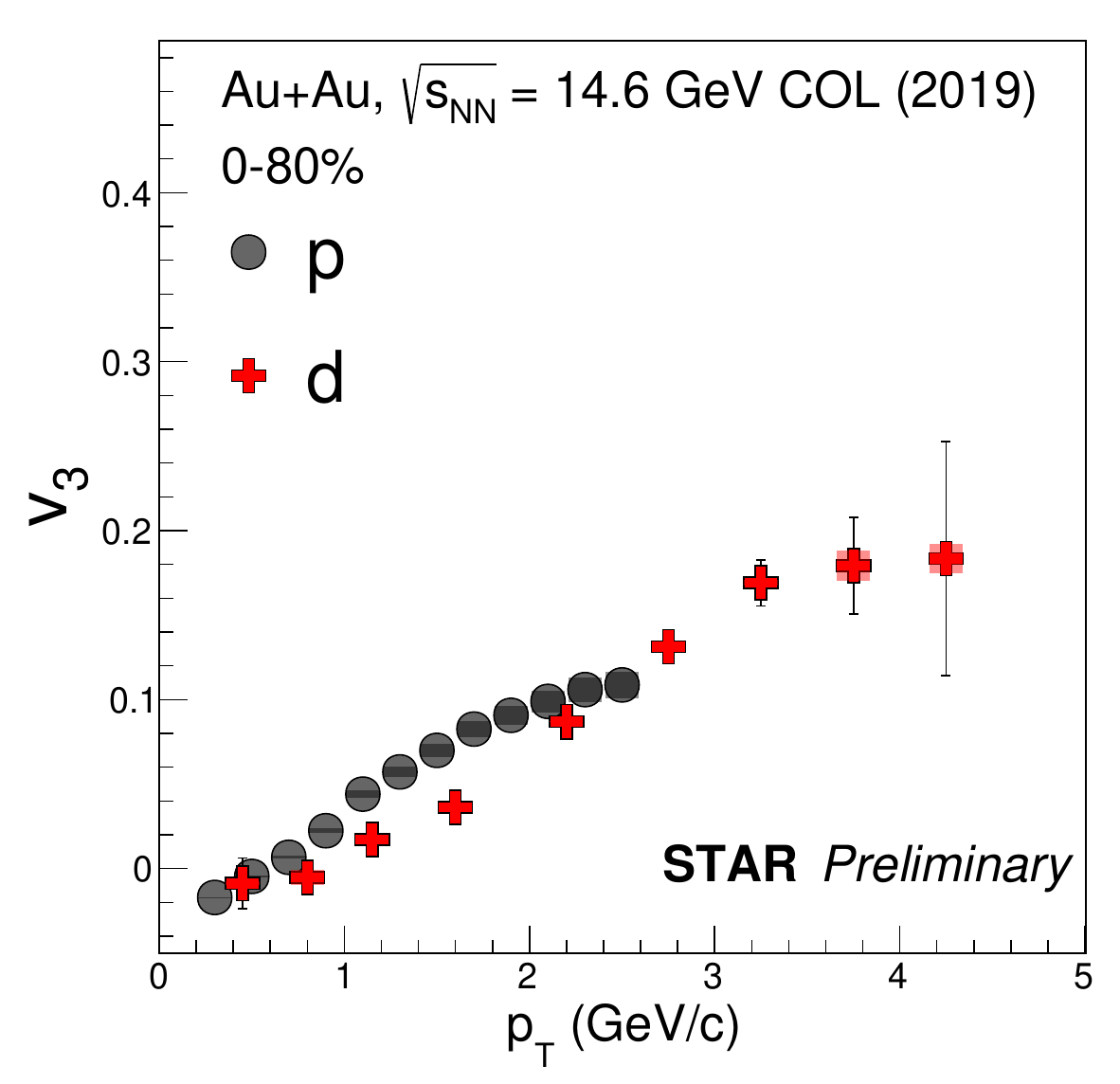}}
    \subfigure{\includegraphics[width=0.24\textwidth]{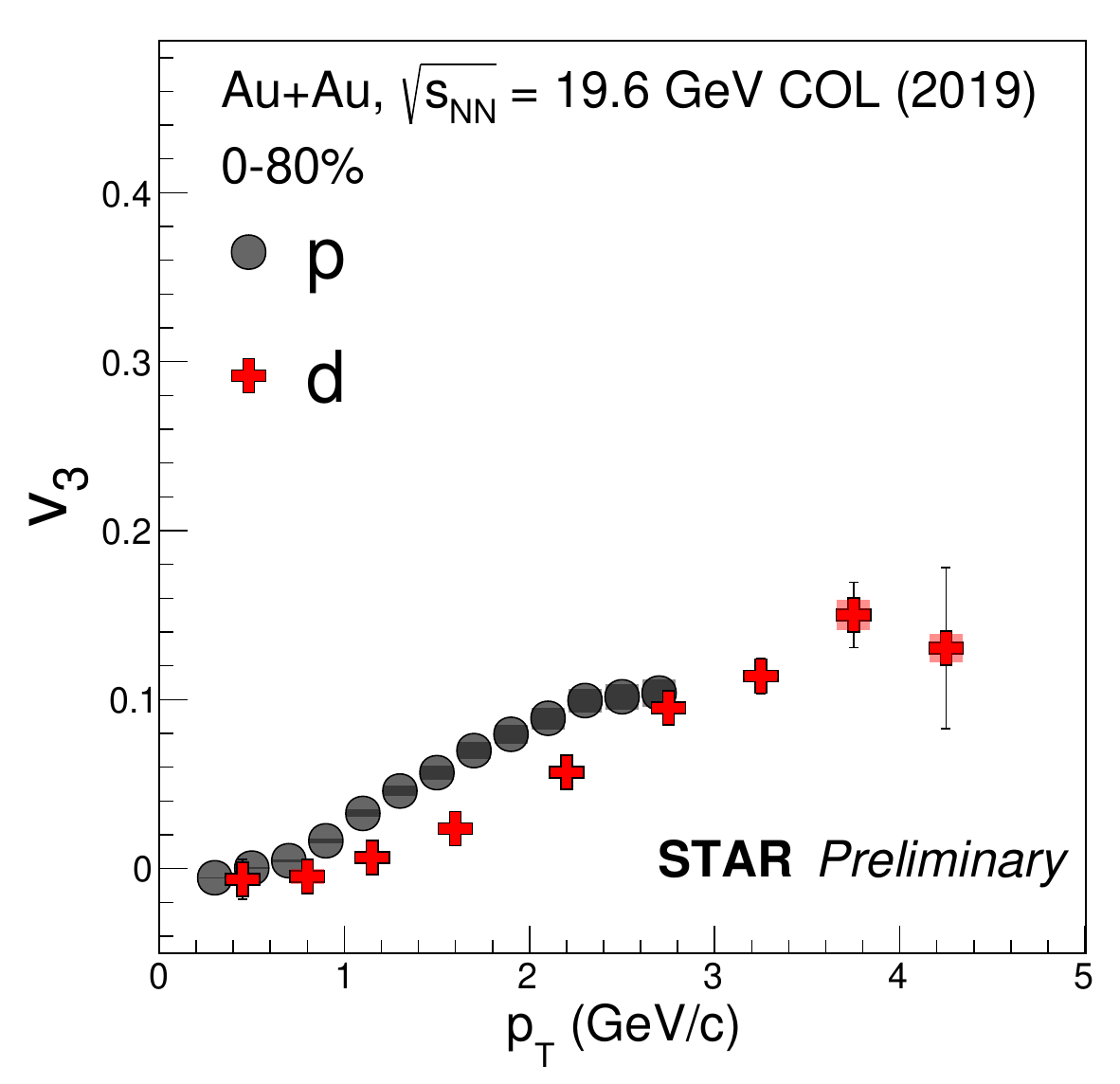}}
     \subfigure{\includegraphics[width=0.24\textwidth]{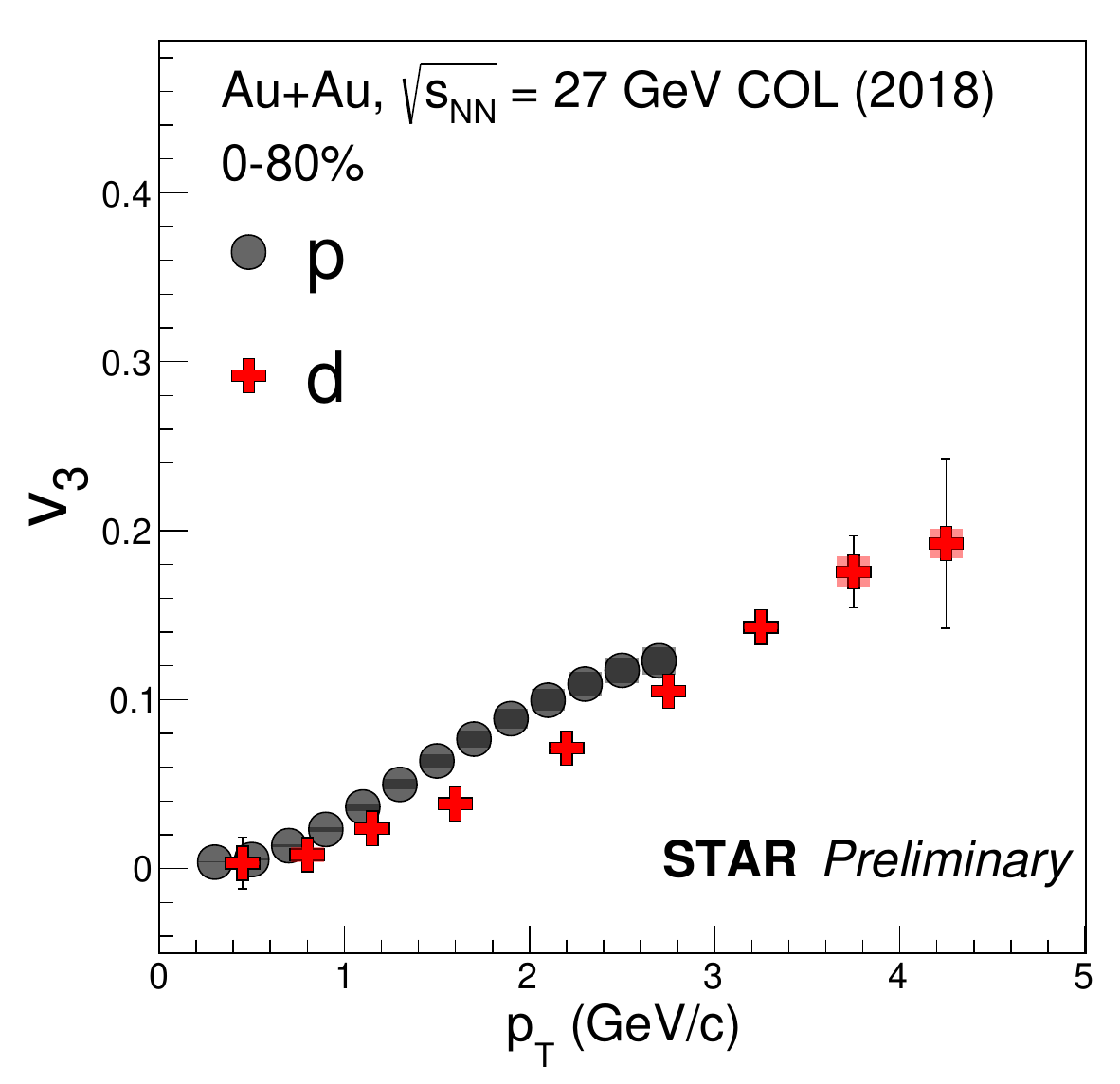}}
    \subfigure{\includegraphics[width=0.24\textwidth]{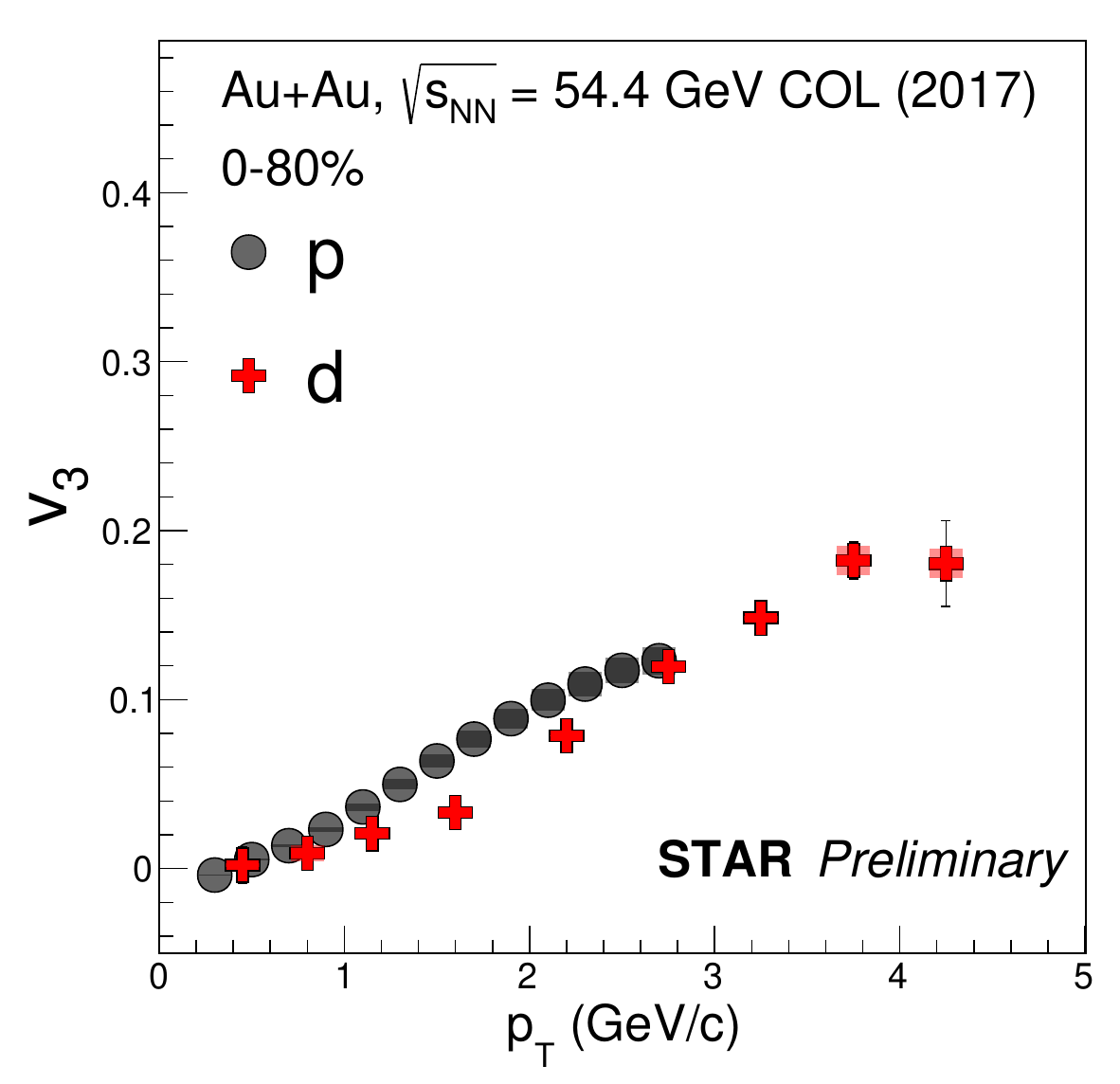}}
    \caption{$v_2(p_T)$ (top panel) and $v_3(p_T)$ (bottom panel) of $p$, $d$, $t$, and $^3\text{He}$ in minimum bias Au+Au collisions at $\sqrt{s_{NN}}$ = 14.6, 19.6, 27, and 54.4 GeV. Vertical lines and shaded area at each marker represent statistical and systematic uncertainties, respectively.}
    \label{fig:flow}
\end{figure}

Figure \ref{fig:flow} shows $v_2$ and $v_3$ of $p$, $d$, $t$, and $^3$He as a function of $p_T$ in 0-80\% centrality interval in Au+Au collisions at $\sqrt{s_{NN}}$ = 14.6, 19.6, 27, and 54.4 GeV. In the measured $p_T$ range, a monotonous increase in $v_2$ and $v_3$ of light nuclei with $p_T$ is observed for all center-of-mass energies. Mass ordering of $v_2$ and $v_3$ is also observed at low $p_T$. 

\subsection{Centrality dependence of $\bm{v_{2}(p_T)}$ of light nuclei}
\label{cent}
\begin{figure}[H]
    \centering
    \subfigure{\includegraphics[width=0.24\textwidth]{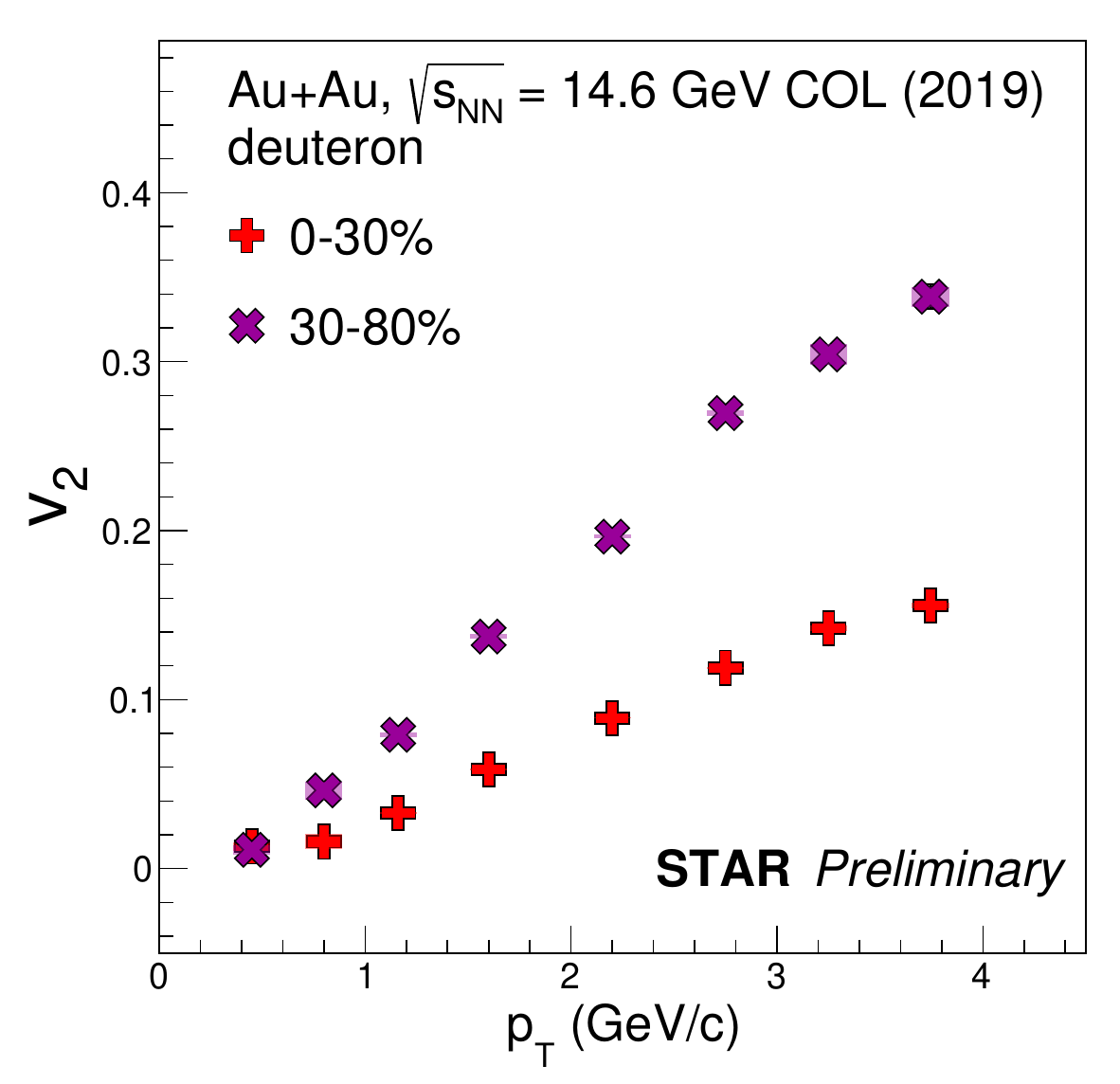}}
    \subfigure{\includegraphics[width=0.24\textwidth]{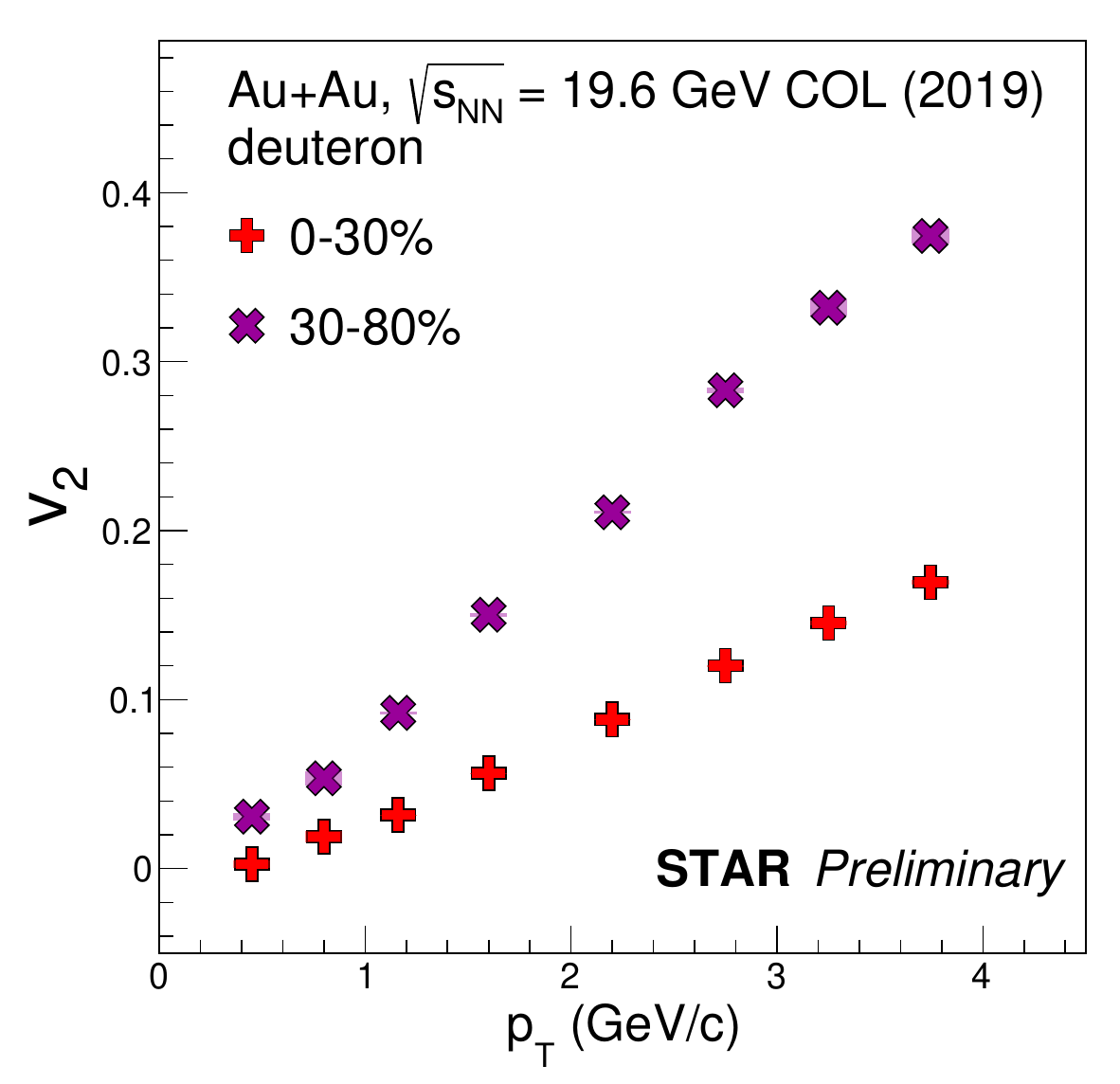}}
     \subfigure{\includegraphics[width=0.24\textwidth]{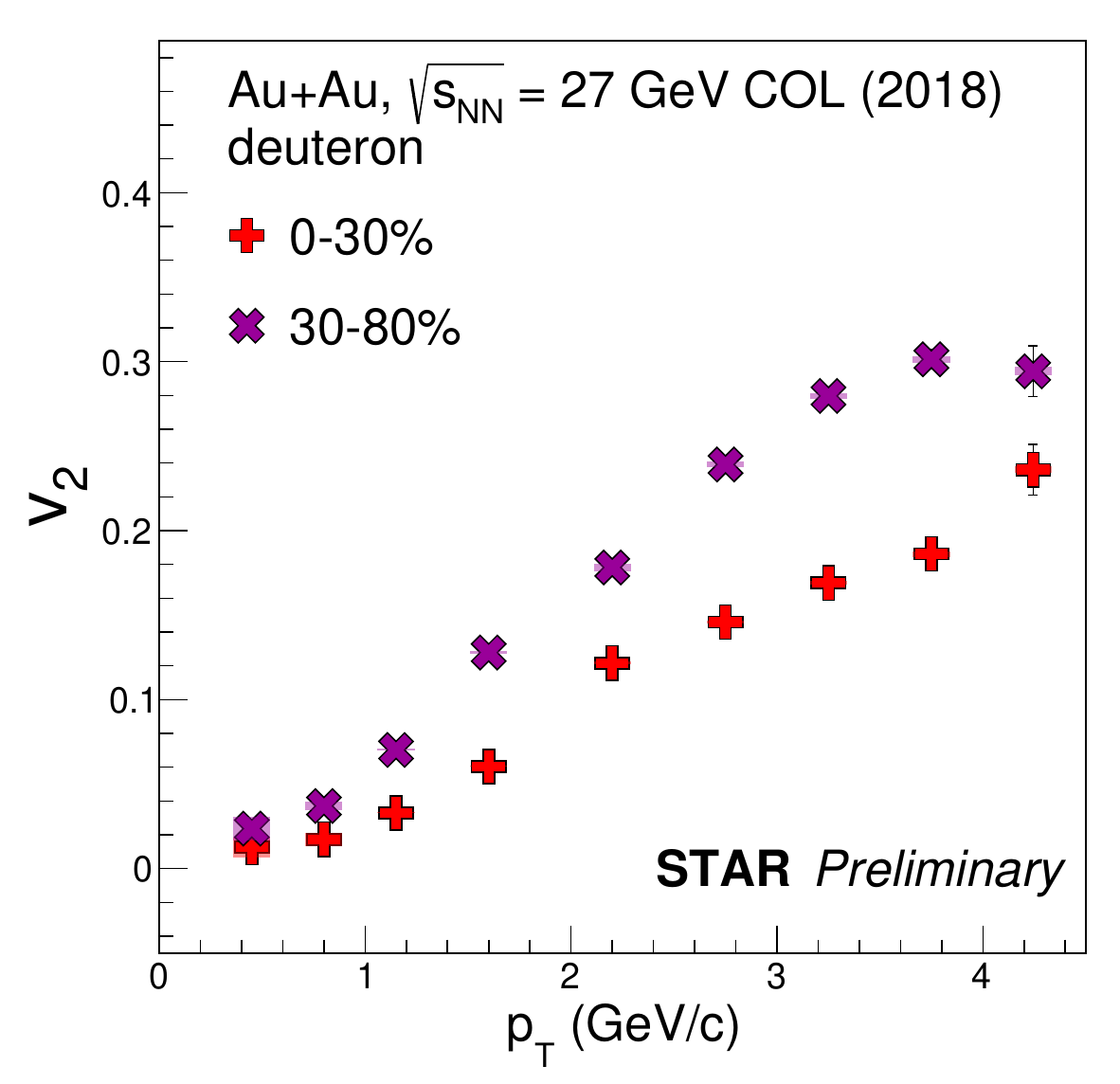}}
    \subfigure{\includegraphics[width=0.24\textwidth]{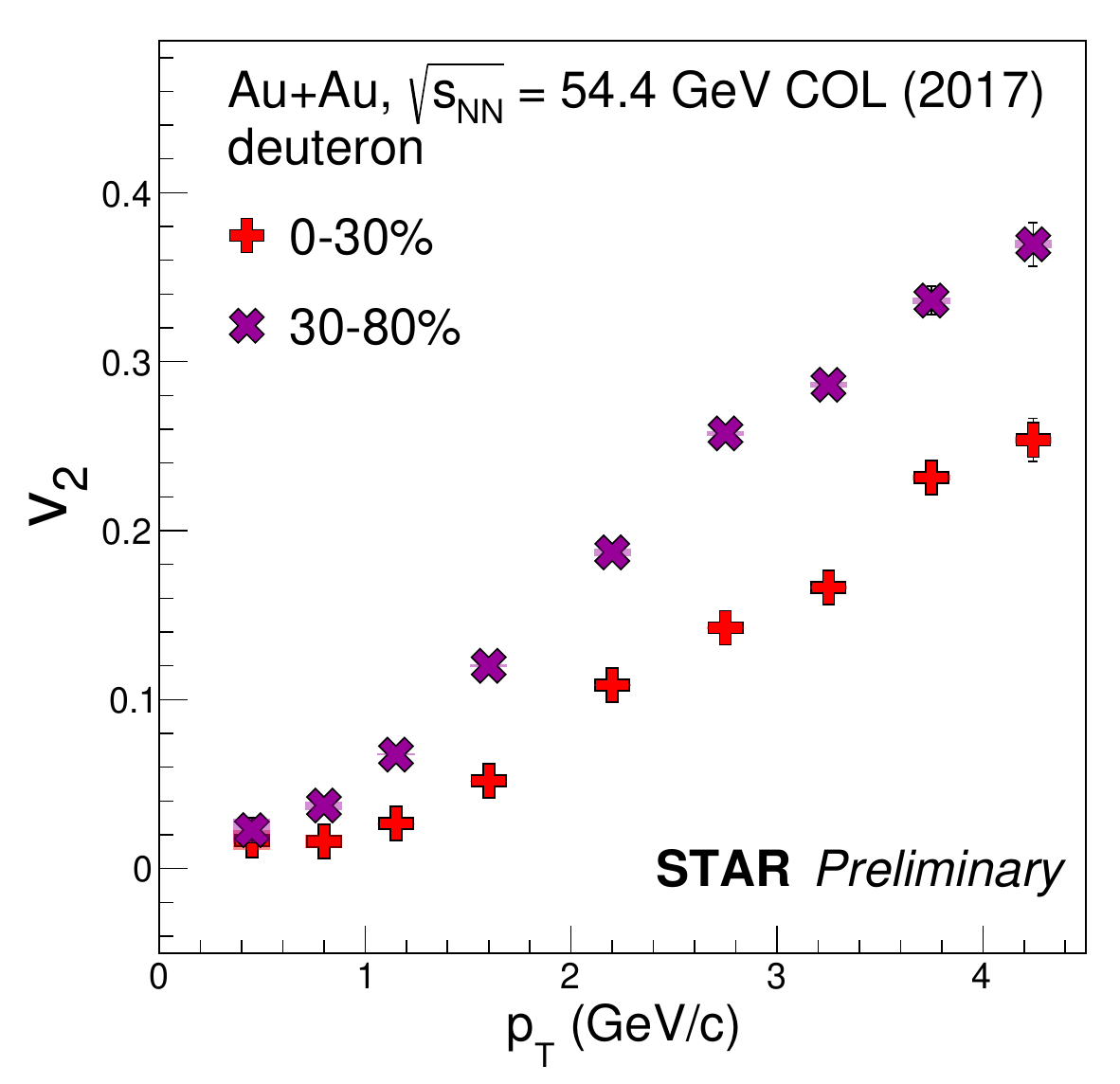}}
    \caption{$v_2(p_T)$ of $d$ measured in 0-30\% and 30-80\% centrality intervals in Au+Au collisions at $\sqrt{s_{NN}}$ = 14.6, 19.6, 27, and 54.4 GeV. Vertical lines and shaded bands at each marker represent statistical and systematic uncertainties, respectively.}
    \label{fig:cent}
\end{figure}

Figure \ref{fig:cent} shows the centrality dependence of $v_2(p_T)$ of $d$ in Au+Au collisions at $\sqrt{s_{NN}}$ = 14.6, 19.6, 27, and 54.4 GeV. $v_2(p_T)$ of $d$ is measured in 0-30\% and 30-80\% centrality ranges in Au+Au collisions. A clear centrality dependence of $v_2(p_T)$ is observed where peripheral collisions have higher $v_2$ values compared to central collisions. This observation can be explained by the fact that peripheral collisions have higher initial spatial anisotropy compared to central collisions, resulting in a higher $v_2$ value.

\subsection{Mass number scaling}
\label{scaling}
\begin{figure}
    \centering
    \subfigure{\includegraphics[width=0.24\textwidth]{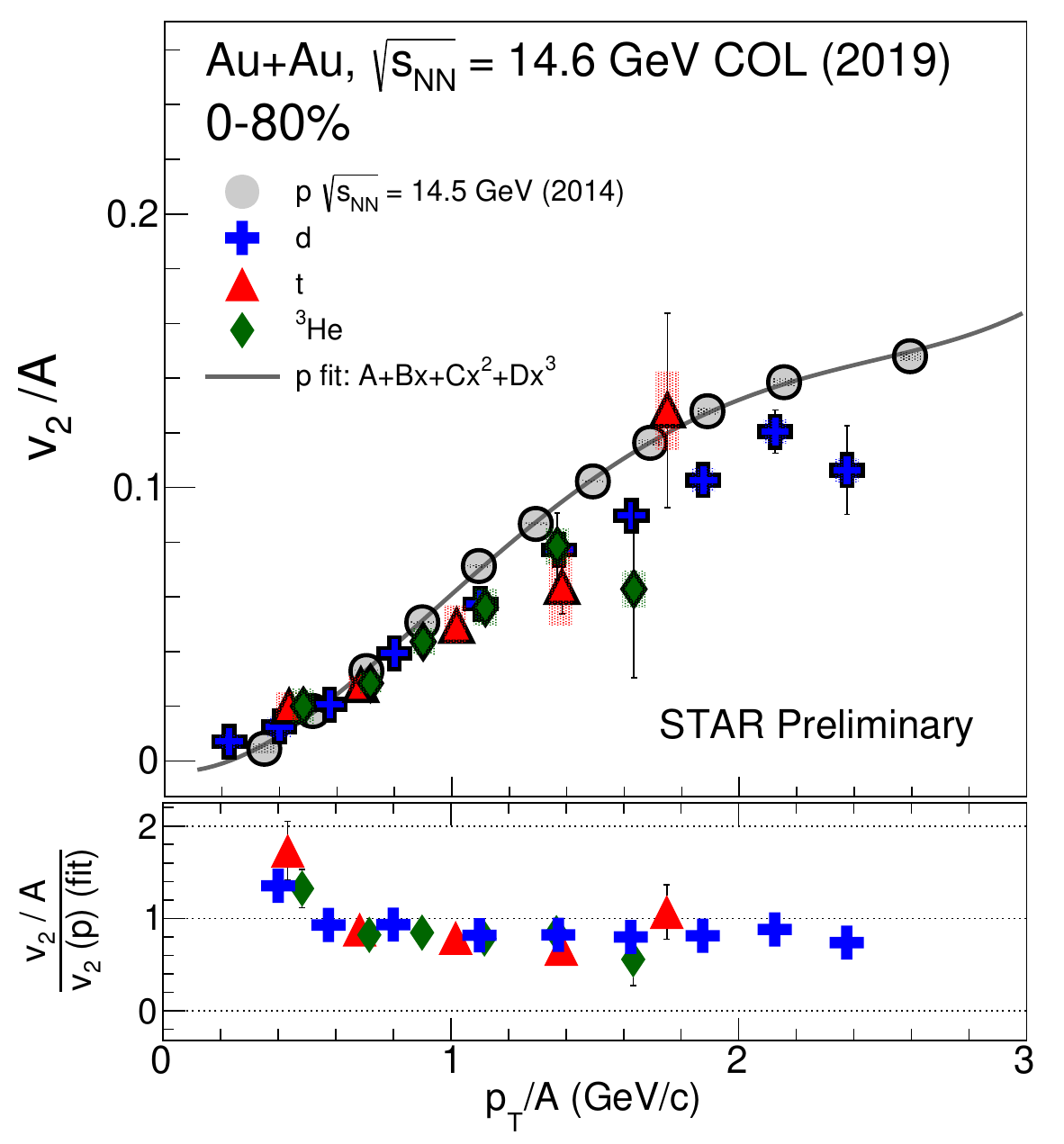}}
    \subfigure{\includegraphics[width=0.24\textwidth]{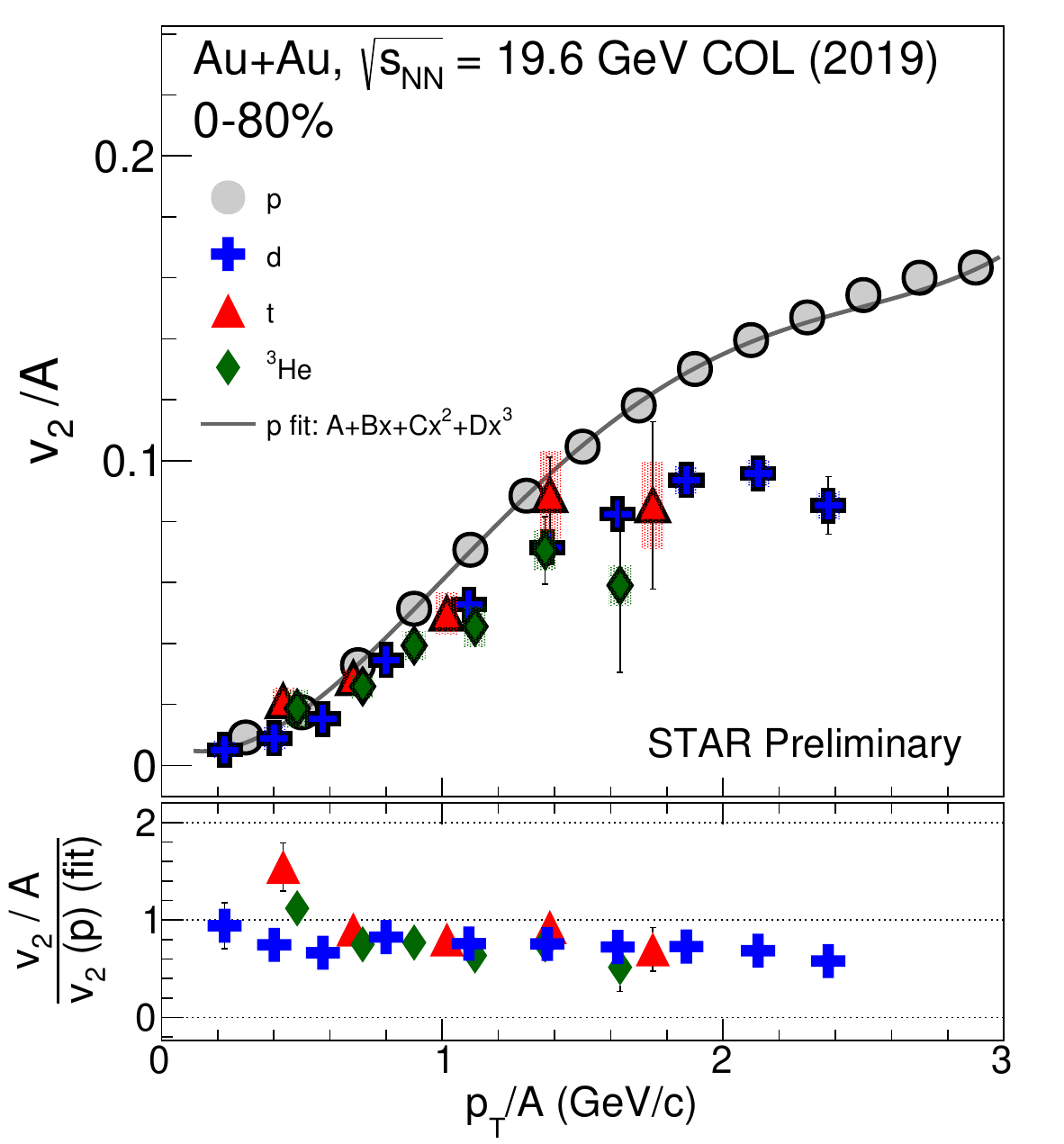}}
     \subfigure{\includegraphics[width=0.24\textwidth]{images/up_mns_19p6_v2.pdf}}
    \subfigure{\includegraphics[width=0.24\textwidth]{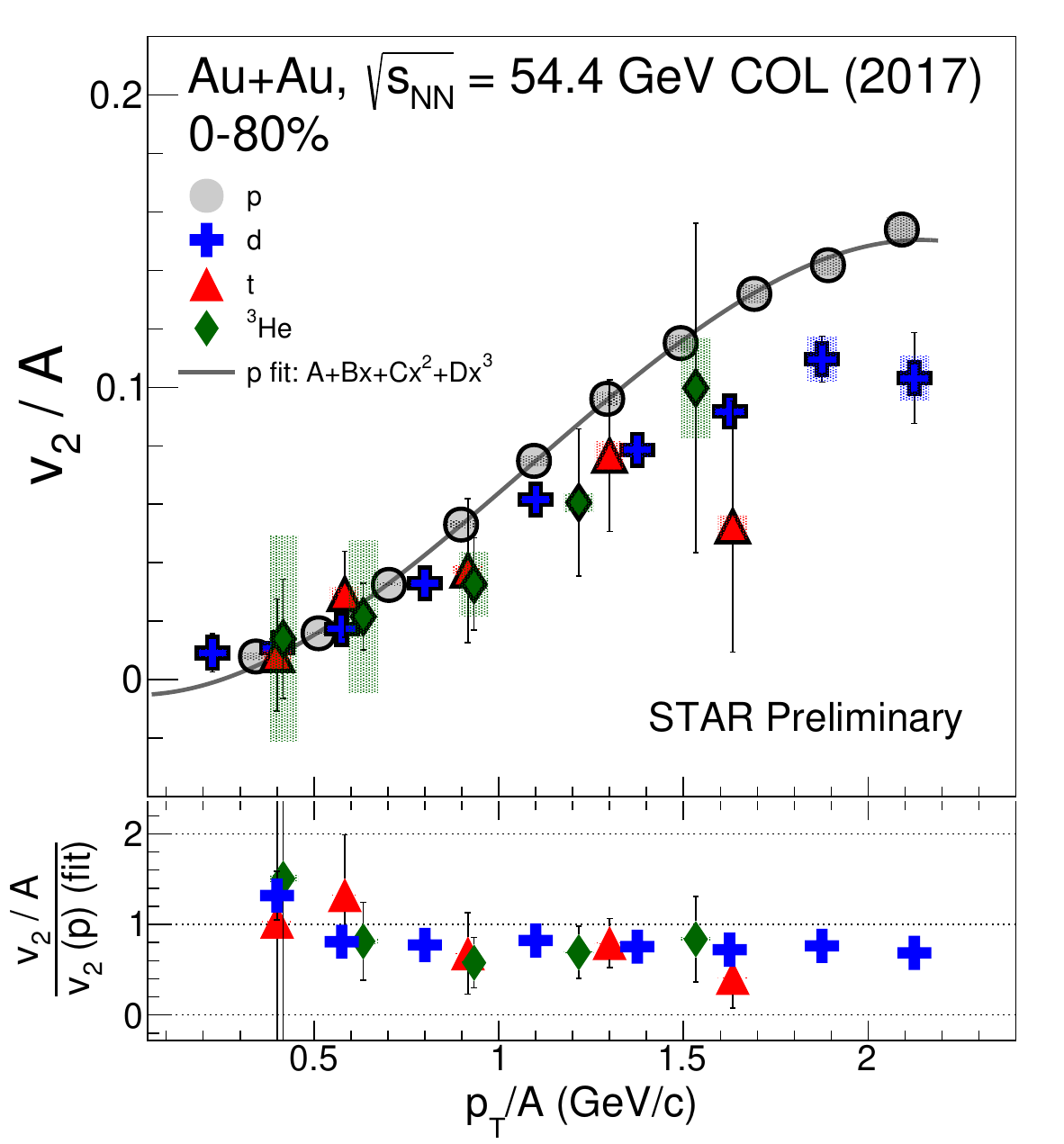}} \newline
    \subfigure{\includegraphics[width=0.24\textwidth]{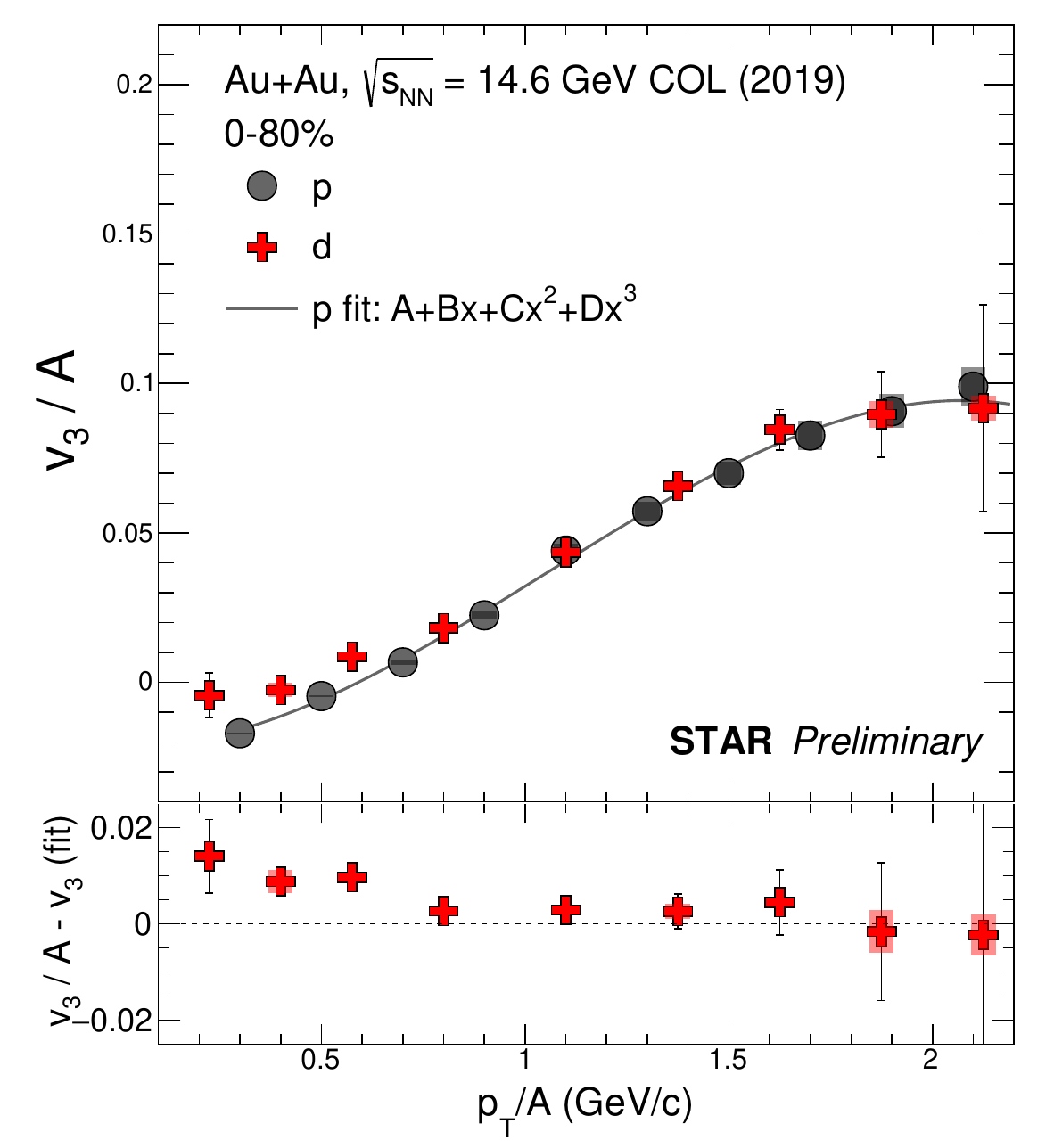}}
    \subfigure{\includegraphics[width=0.24\textwidth]{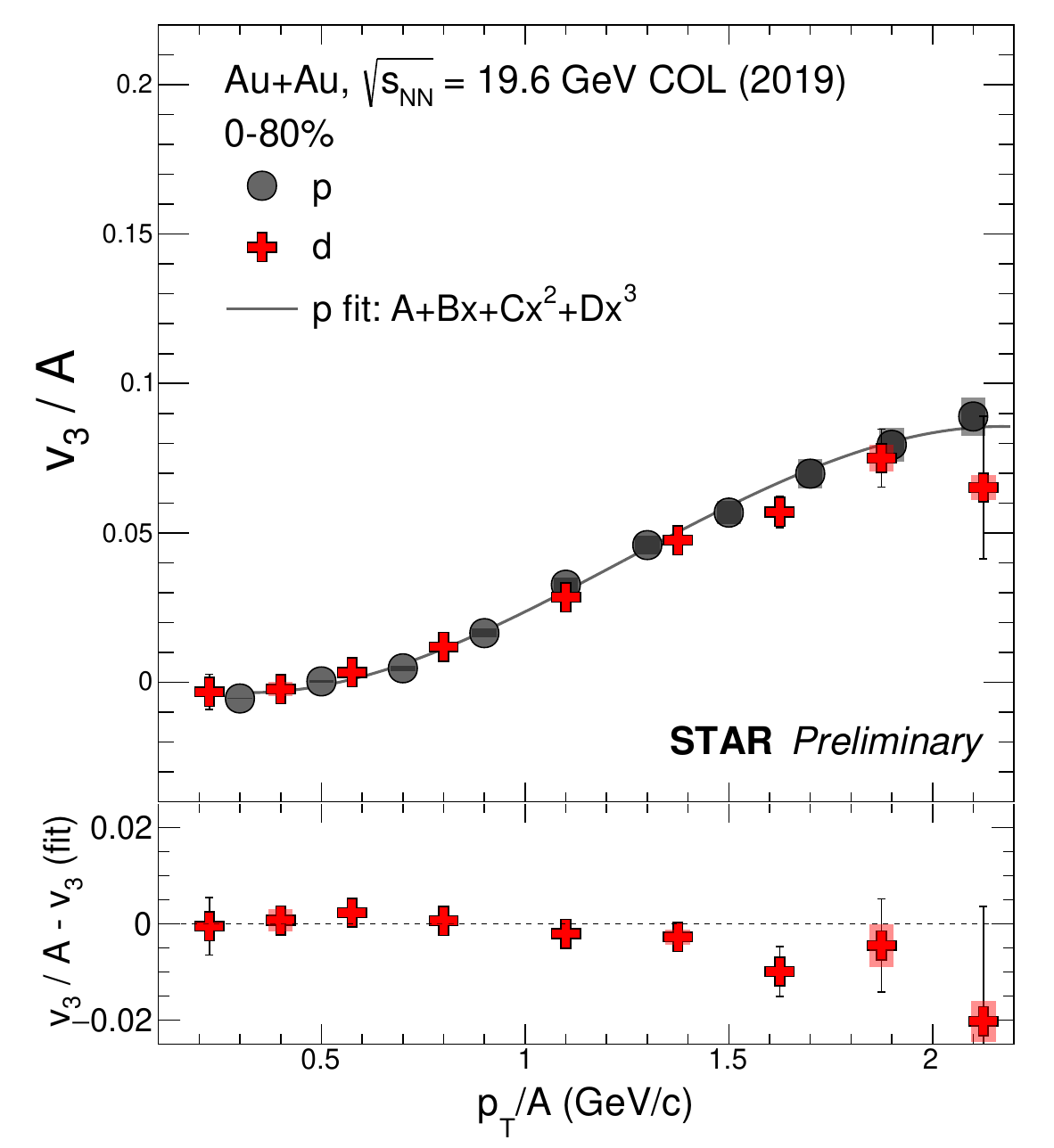}}
     \subfigure{\includegraphics[width=0.24\textwidth]{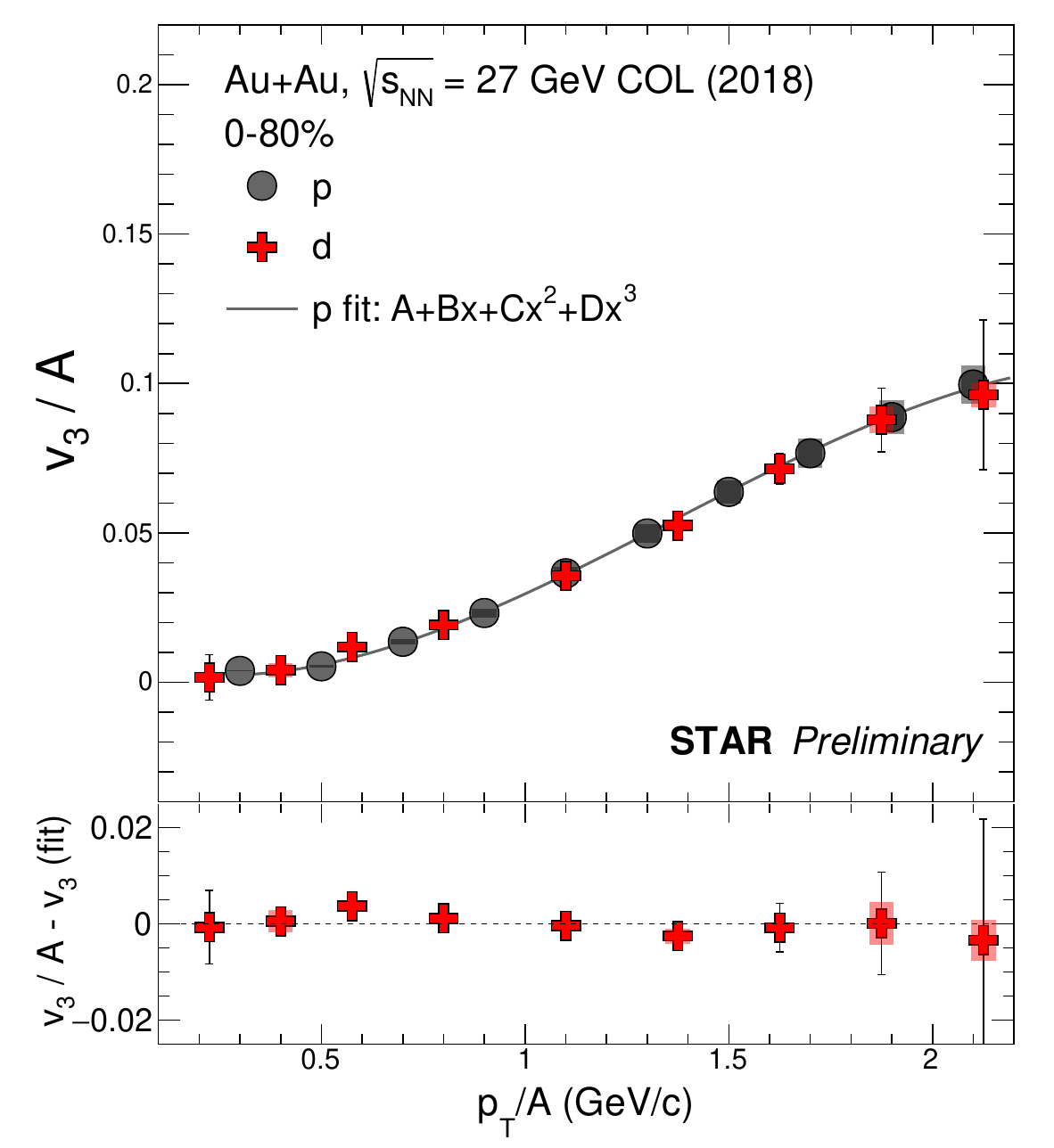}}
    \subfigure{\includegraphics[width=0.24\textwidth]{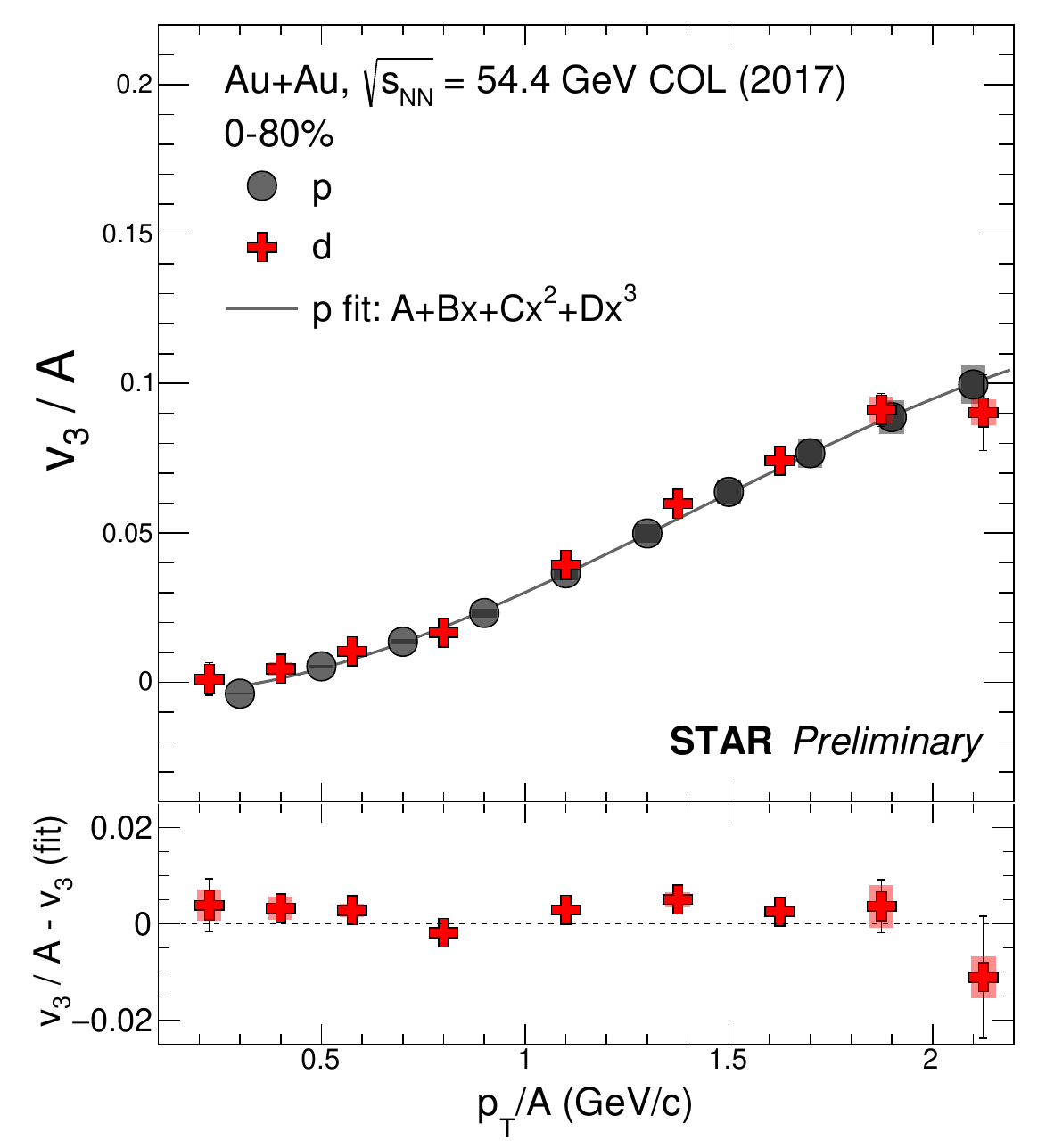}}
    \caption{Mass number scaling of $v_2$ of $p$, $d$, $t$, and $^3$He (top panel) and $v_3$ of $p$ and $d$ (bottom panel) as a function of $p_T/A$ in minimum bias Au+Au collisions at $\sqrt{s_{NN}}$ = 14.6, 19.6, 27, and 54.4 GeV. Gray solid lines correspond to third order polynomial fits to $v_2$ and $v_3$ of $p$. The ratios of [$v_2/A$]/fit for $d$, $t$, and $^3$He and the differences of [$v_3/A$]-fit for $d$ are shown for each collision energy. Vertical lines and shaded bands at each marker represent statistical and systematic uncertainties, respectively.
}
    \label{fig:scaling}
\end{figure}

Figure \ref{fig:scaling} shows the comparison of $v_{2}/A$ and $v_{3}/A$ of light nuclei as a function $p_T/A$, where $A$ is the mass number of the corresponding nuclei, with $v_2/A$ of proton ($A=1$ for proton). The aim of this study is to compare $v_{2}$ and $v_3$ of light nuclei with the expectation of mass number scaling from the coalescence picture. According to the coalescence model, assuming $v_2$ ($v_3$) of proton and neutron is identical, for a light nuclei species $N$ with mass number $A$, it is expected that, $v^N_{2(3)}(p_T) \approx Av^p_{2(3)}(p_T/A)$, where $v^p_{2(3)}$ is the elliptic (triangular) flow of proton \cite{MSN1,MSN2,MSN3}. 
It is observed that $v_2$ of $d$, $t$, and $^3$He deviates from mass number scaling by 20-30\% for the measured center-of-mass energies. However, $v_3$ of $d$ is observed to follow mass number scaling within 10\% for the measured center-of-mass energies. 

\subsection{Comparison with AMPT and coalescence calculations}
\label{model_coal}
\begin{figure}
    \centering
    \subfigure{\includegraphics[width=0.24\textwidth]{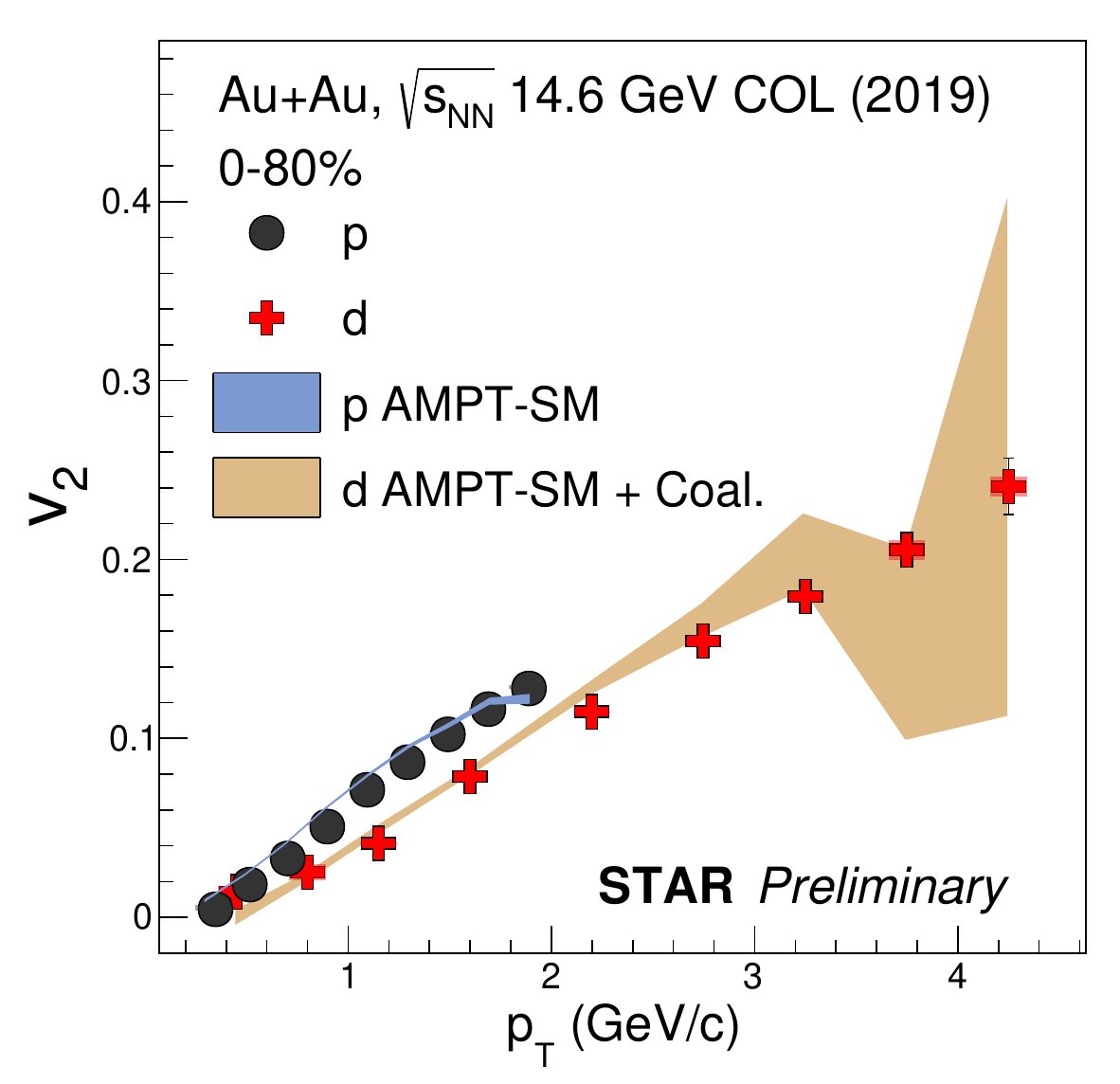}}
    \subfigure{\includegraphics[width=0.24\textwidth]{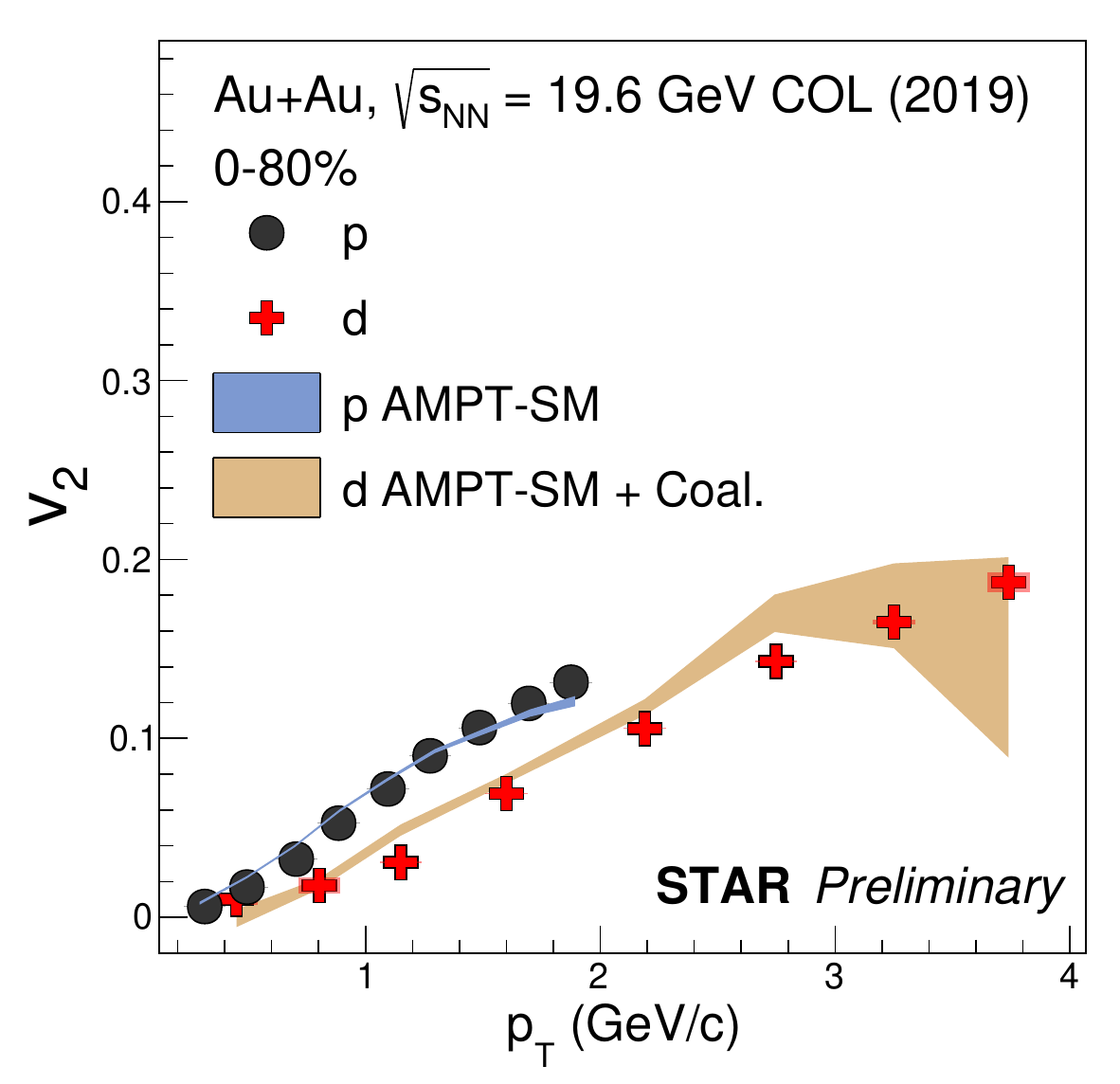}}
     \subfigure{\includegraphics[width=0.24\textwidth]{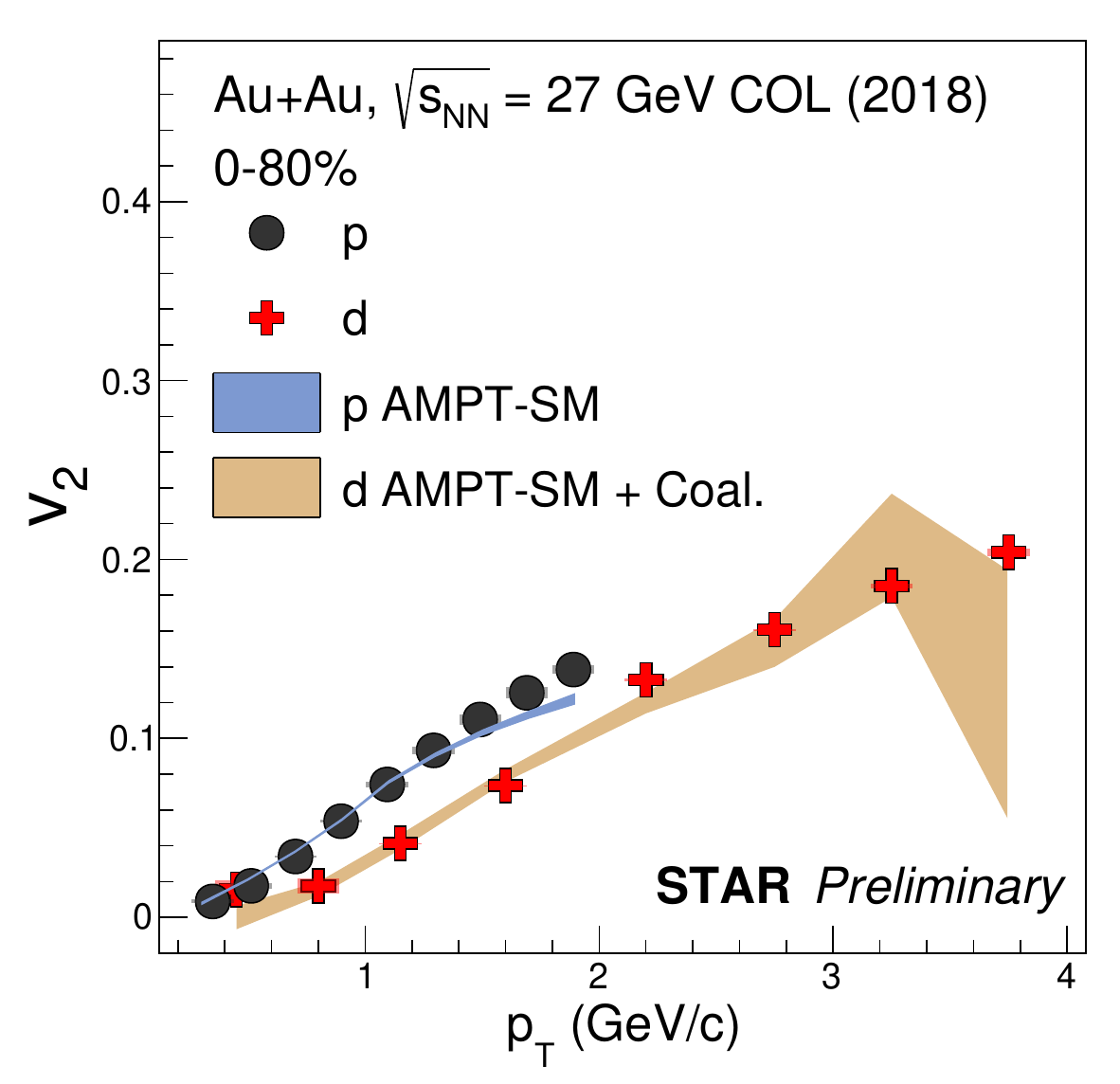}}
    \subfigure{\includegraphics[width=0.24\textwidth]{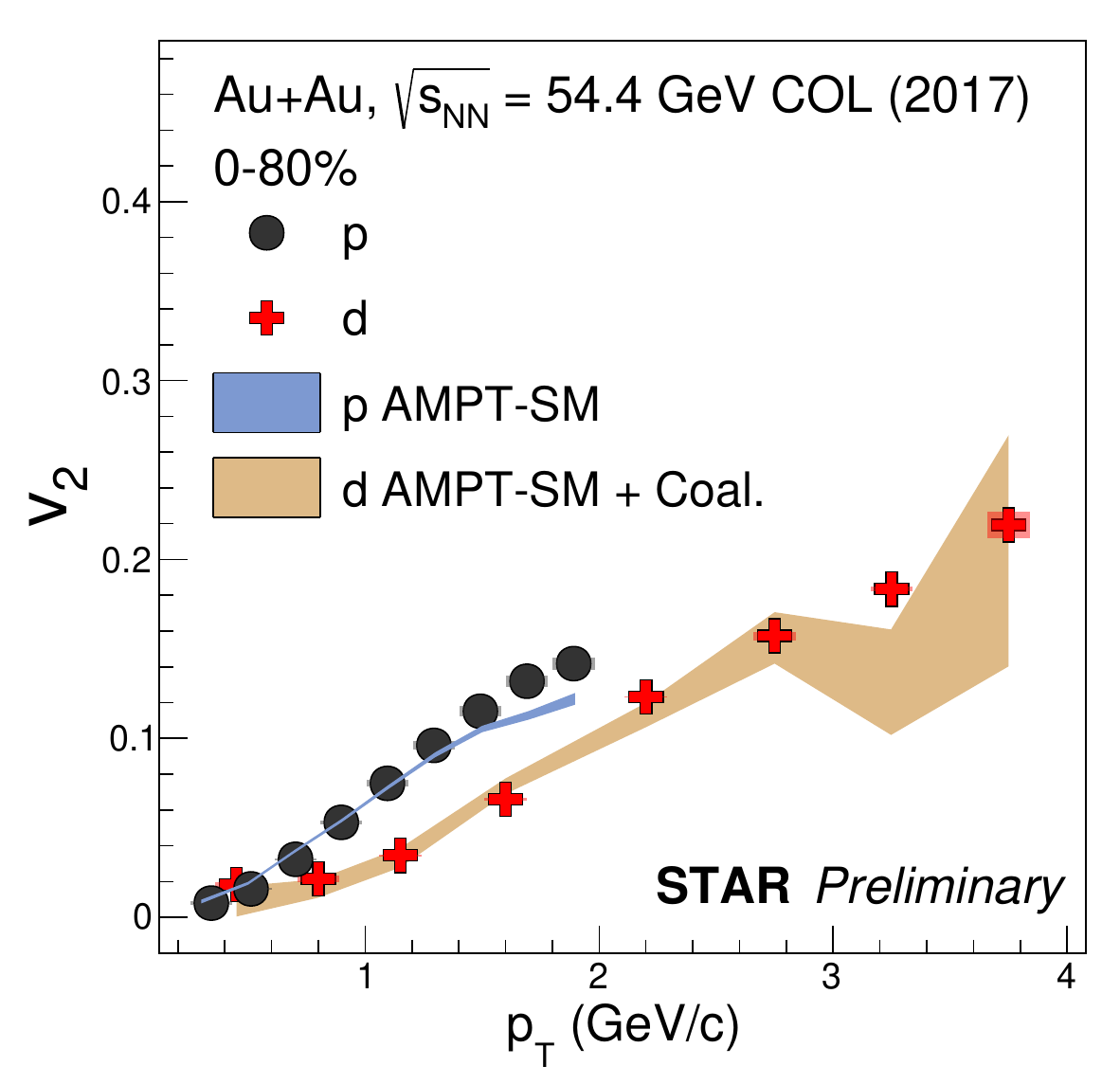}} 
    \newline
    \subfigure{\includegraphics[width=0.24\textwidth]{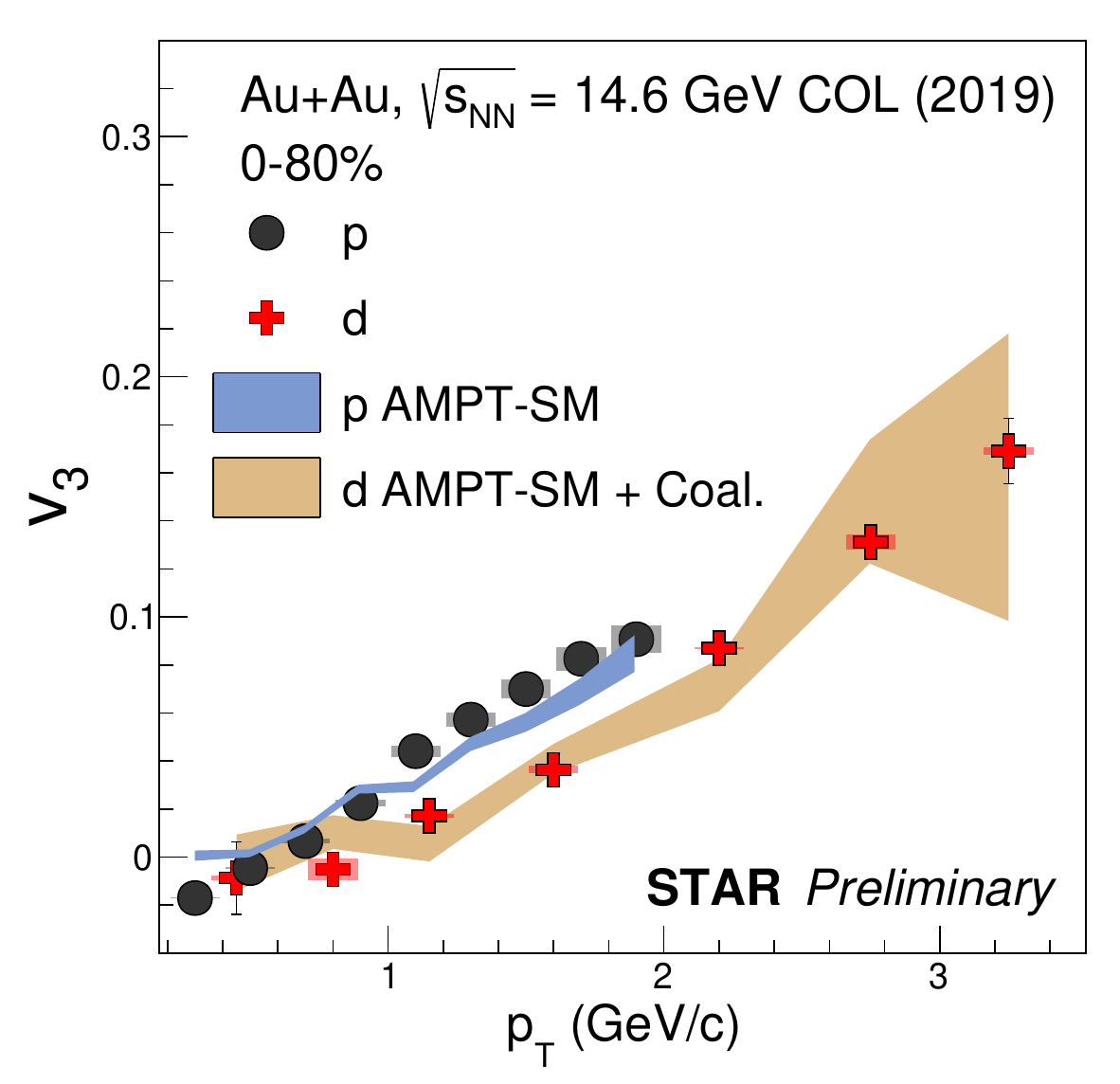}}
    \subfigure{\includegraphics[width=0.24\textwidth]{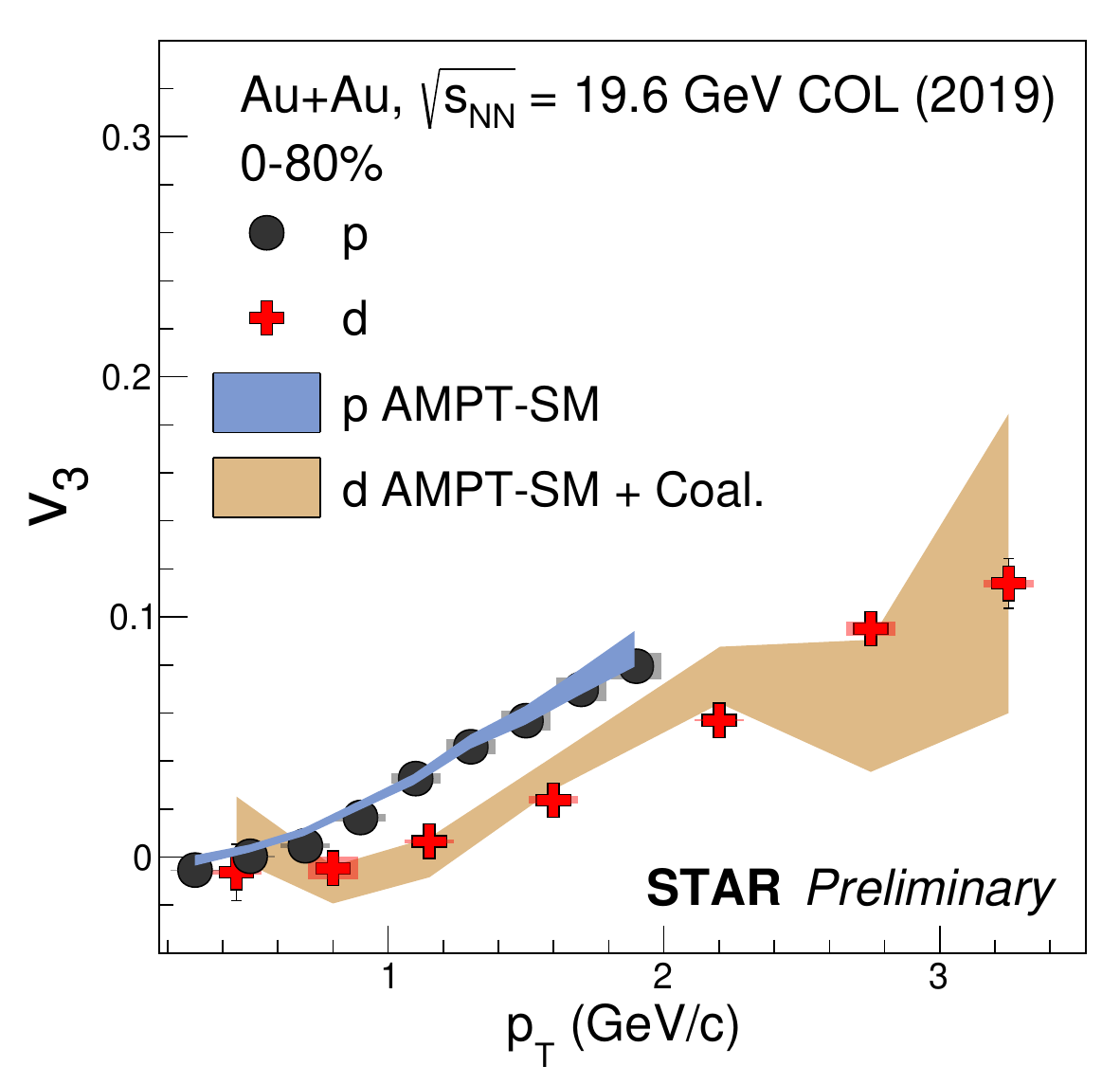}}
     \subfigure{\includegraphics[width=0.24\textwidth]{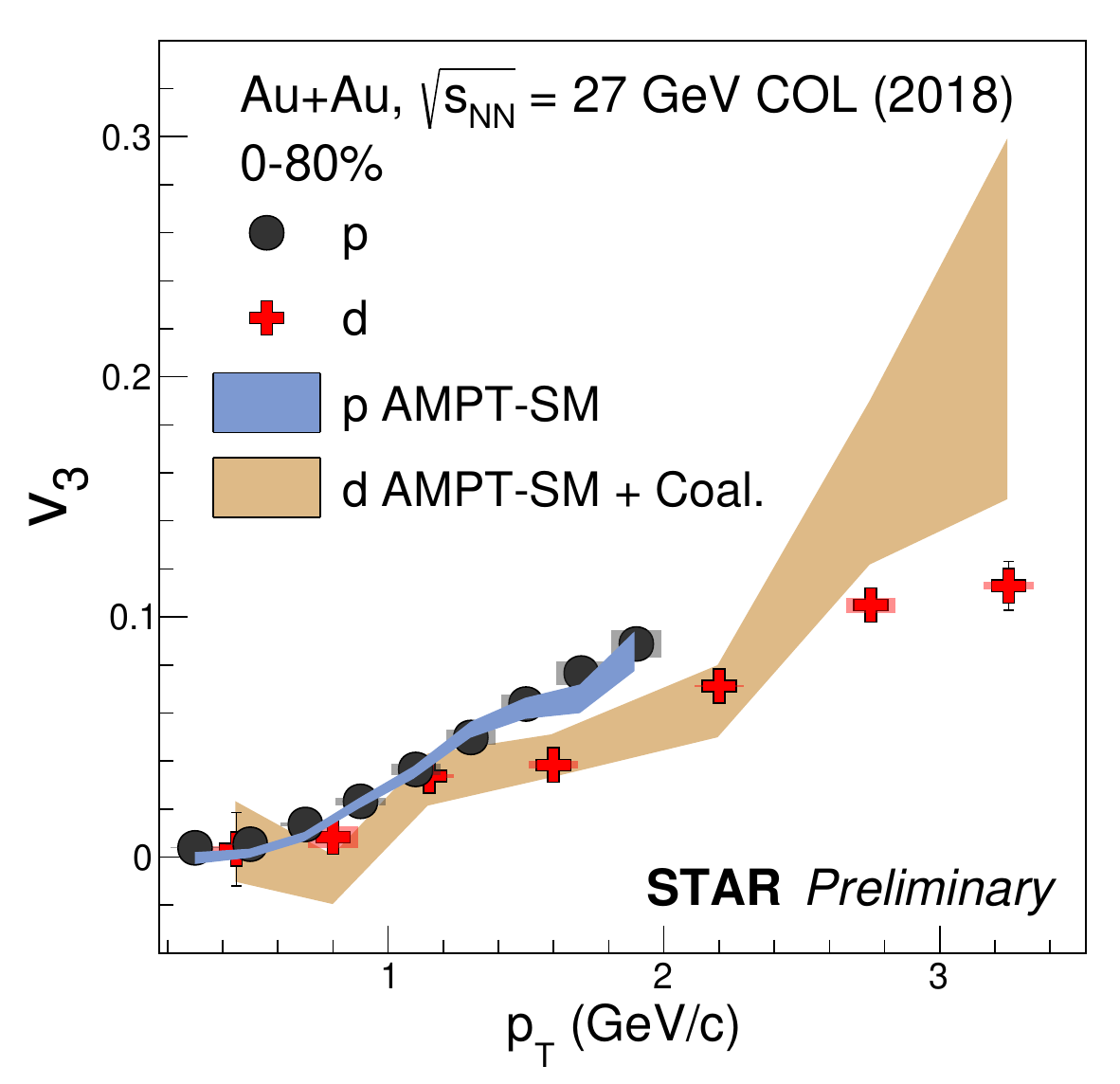}}
    \subfigure{\includegraphics[width=0.24\textwidth]{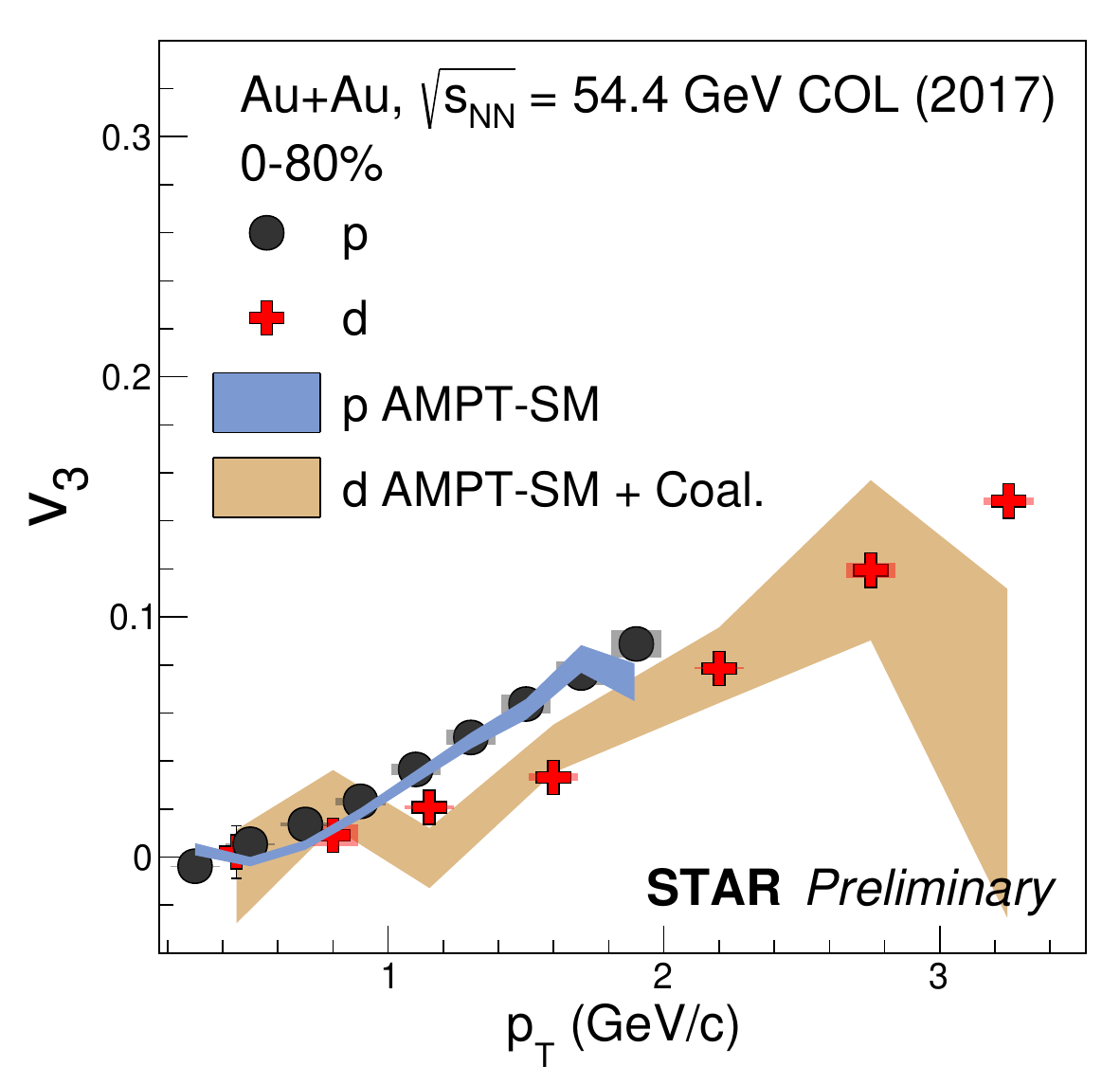}}
    \caption{$v_2(p_T)$ (top panel) and $v_3(p_T)$ (bottom panel) of $p$ and $d$ compared with the results of AMPT+coalescence calculations (solid bands).}
    \label{fig:coal}
\end{figure}

To further test the hypothesis of light nuclei production from nucleon coalescence, string melting version of A Multi Phase Transport (AMPT, version ampt-v1.26t9b-v2.26t9b) model \cite{ampt} in conjunction with a dynamic coalescence model is used to get a theoretical estimate of $v_{2}$ and $v_{3}$ of deuteron. In this model, the probability of coalescence is determined by the superposition of the Wigner function of the deuterons and nucleon phase-space distribution at freeze-out obtained from AMPT \cite{coal}.
Figure \ref{fig:coal} shows the comparison of $v_2$ and $v_3$ of $d$ with the results from AMPT+Coalescence calculations. We have also compared $v_2$ and $v_3$ of $p$ with AMPT model calculations. The data and model are observed to agree within uncertainties in the measured center-of-mass energies and $p_T$ ranges. The agreement of $v_2$ and $v_3$ of $d$ with the calculations from AMPT+Coalescence model indicate that final-state nucleon coalescence might be the dominant production mechanism of light nuclei. 

\section{Summary}
\label{summary}
In summary, we have reported $v_2$ of $d$, $t$, and $^3\text{He}$ and $v_3$ of $p$ and $d$ in Au+Au collisions at $\sqrt{s_{NN}}$ = 14.6, 19.6, 27, and 54.4 GeV. A monotonic rise in light nuclei $v_2$ and $v_3$ with $p_T$ in the measured energy and $p_T$ range is observed. Mass ordering of $v_2$ and $v_3$ of light nuclei at low $p_T$ is observed. $v_2$ of $d$ is observed to show a strong centrality dependence being higher for peripheral collisions compared to central collisions. $v_2$ of light nuclei is found to deviate by 20-30\% from mass number scaling in the measured center-of-mass energies whereas $v_3$ of $d$ is observed to be in agreement with mass number scaling within 10\%. In addition, we also observe that $v_2$ and $v_3$ of $d$ are well described by the model calculations using AMPT+Coalescence, within uncertainties. These observations suggest that the final-state coalescence of nucleons might be the dominant mechanism of light nuclei production in heavy-ion collisions.

\end{document}